\documentclass[pre,twocolumn]{revtex4-1}

\usepackage{graphicx}
\usepackage{dcolumn}
\usepackage{bm}

\usepackage[utf8]{inputenc}
\usepackage[T1]{fontenc}
\usepackage{mathptmx}
\usepackage{bm}
\usepackage{pdfrender}

\usepackage{amsmath}
\usepackage{mathtools}
\usepackage{amssymb}
\usepackage{tikz}

\usetikzlibrary{matrix}
\usepackage{amsmath, tikz-cd}
\usetikzlibrary{arrows.meta}

\usepackage{pst-node}
\usepackage{verbatim}   
\usepackage{color}      
\usepackage{subfigure}  
\usepackage{hyperref}   

\usepackage{braket} 

\newcommand{\bnabla}{\mbox{\boldmath $\nabla$}}

\newcommand{\ba}{\begin{eqnarray}}
\newcommand{\ea}{\end{eqnarray}}
\newcommand{\be}{\begin{equation}}
\newcommand{\ee}{\end{equation}}

\begin{document}

\title{Active oscillator: recurrence relation approach} 

\author{Derek Frydel}
 \email{dfrydel@gmail.com}
\affiliation{Department of Chemistry, Universidad Técnica Federico Santa María, Campus San Joaquin, 7820275, Santiago, Chile}

\date{\today}

\begin{abstract}
The present work analyzes stationary distributions of active Brownian particles in a harmonic trap. 
Generally, obtaining stationary distributions for this system is non-trivial, and up to date no exact 
expressions are available. In this work, we develop and explore a method based on a transformation 
of the Fokker-Planck equation into a recurrence relation for generating moments of a distribution. 
The method, therefore, offers an analytically tractable approach, an alternative to numerical simulations, 
in a situation where more direct analytical approaches fail. Although the current work focuses on the 
active Brownian particle model, the method is general and valid for any type of active dynamics and 
any system dimension.
\end{abstract}

\pacs{
}

\maketitle

\section{Introduction}

Recently, Ref. \cite{Frydel23b} developed an approach for treating run-and-tumble particles (RTP) in a harmonic 
trap in three-dimensions.  The approach was based on a transformation of a Fokker-Planck equation into a 
recurrence relation that can generate moments of a stationary distribution.  While an exact distribution is 
unavailable, the generated moments have a straightforward algebraic form.  (It is noted that in one- and 
two-dimensions, stationary distributions of the RTP model have a simple beta functional form and in this case 
no alternative treatment is required \cite{Cates08,Cates09,Dhar19,Frydel22c}).  

In this work, we extend the method developed in Ref. \cite{Frydel23b} to the active Brownian particle model (ABP), 
in which case, no exact solutions are available in any dimensions.  The resulting recurrence relation, obtained from 
the corresponding Fokker-Planck equation, has a more complex structure than the analogous relation for RTP particles.  
Consequently, it cannot be reduced to a simpler form.  Nonetheless, the relation can easily be used to generate an 
arbitrary number of moments almost instantaneously.   As for other cases, the moments have a simple algebraic form.  

To simplify the procedure, we take advantage of certain features of a harmonic potential.  First, we apply 
our procedure to a harmonic potential in a single direction, $u = \frac{1}{2} Kx^2$, but that is embedded in 
a higher dimensional space.  This allows us to represent the problem as effectively one-dimensional.  The 
system dimension appears as a parameter in the resulting differential equation.  It turns out that the moments 
of a stationary distribution of this "effectively" one-dimensional system can be related to the moments of a 
stationary distribution for an isotropic potential $u = \frac{1}{2} K r^2$ via a very simple relation.

The second simplification results from the neglect of thermal fluctuations.  The procedure outlined above is 
carried out for a system at zero temperature.  Thermal fluctuations are incorporated afterwards by taking 
advantage of another feature specific to harmonic potentials which is that thermal fluctuations are mathematically 
incorporated using a convolution construction \cite{Frydel22c,Scher22,Frydel23}.  (This is a consequence of 
the fact that for a harmonic potential thermal fluctuations and active motion remain independent.)  The convolution 
construction permits us to express moments with thermal fluctuations as an expansion of moments at zero 
temperature.   

The analysis developed in this work extends analytical tractability of a system that otherwise would have to be 
treated using numerical simulations based on the integration of a Langevin equation.  This analytical tractability 
reveals certain relations, discussed later in the body of the work, that otherwise could not be determined based on
simulations.  The analysis, furthermore, underscores intrinsic complexity of the ABP model.  The fact that the 
moments have to be generated from a recurrence relation and there is no general expression of an arbitrary 
moment indicates an underlying mathematical complexity of stationary distributions.

Active particles in a harmonic trap has garnered interest both as a minimalistic theoretical model and as an 
experimentally realizable system.  Experimental realizations of active particles in a harmonic trap can be carried
out using acoustic \cite{Brady16} or optical \cite{Lowen22} tweezers.  Least challenging active particle model 
from theoretical point of view is  
the active Ornstein-Uhlenbeck particle model (AOUP).  Stationary distributions in this case have a Gaussian 
functional form \cite{Szamel14,Fodor16,Martin21}.  In the case of RTP particles, a stationary distribution in 
one- and two-dimensions corresponds to a beta function \cite{Cates08,Cates09,Dhar19,Frydel22c}.  An extension
 of the RTP model in 1D and zero 
temperature to three discrete swimming velocities was considered in \cite{Basu20}.  A stationary distribution 
in this case is available and corresponds to a hypergeometric function.  The RTP model in two-dimensions 
(2D) with four discrete swimming velocities was investigated in \cite{Scher22}.    A unifying approach to 
theoretical treatment of ABP and AOUP models in a harmonic trap was investigated in \cite{Carpini22}.  
Rare events in the context of active particles in a 
harmonic potential were considered in \cite{Farago22}.  Active particles in a harmonic chains was considered in 
\cite{Gupta21,Kundu21,Basu22}.  Entropy production rate of 
active particles in a harmonic trap was studied in \cite{Dabelow19,Caprini19,Dabelow21,Pruessner21,Frydel22a,Frydel23}.

Articles that are most relevant to the present work are Ref. \cite{Dhar20,Cargalio22}.  Both studies are dedicated 
to ABP model in a harmonic trap in two-dimensions.  The work in \cite{Dhar20} provides a series expansion of 
stationary distributions at a finite temperature and the work in \cite{Cargalio22} analyzes dynamics of ABP particles.  
An important result of Ref. \cite{Cargalio22}  is a moment formula of a stationary distribution, in that work given in 
Eq. (21).  In the current work, we provide a different procedure for calculating moments of a distribution.
Furthermore, our results are general and apply to any dimension and active dynamics.

\section{ABP model: the Fokker-Planck equation}

We start with the general Fokker-Planck equation (FP) for active particles subject to an external force ${\bf F}$ 
(and ignoring thermal fluctuations):  
\be
\dot\rho =  -\bnabla\cdot \left[ \left( {\bf F}  -  v_0 {\bf n} \right) \rho \right] + \hat L\, \rho.  
\label{eq:FP}
\ee
In the above equation, ${\bf n}$ is the unit vector designating orientation of the swimming velocity $v_{swim} = v_0 {\bf n}$, 
in 2D defined as ${\bf n} = (\cos\theta,\sin\theta)$, where $\theta$ is the angle of the orientation.  The evolution 
of the vector ${\bf n}$ is governed by the operator $\hat L$, where the type of evolution determines the type of 
active dynamics.  For the ABP model the vector ${\bf n}$ undergoes diffusion.  To ensure that the total number of 
particles is conserved, $\int_0^{2\pi} d\theta\,\hat L\, \rho = 0$.  

For a harmonic potential in a single direction, $u = \frac{1}{2} Kx^2$, the FP equation for the ABP model, 
under steady-state conditions, becomes 
\be
0 =  \frac{\partial}{\partial x} \left[ \left( \mu K x  -  v_0 \cos\theta \right) \rho \right] 
+  D_r  \left[ \frac{\partial^2 \rho}{\partial\theta^2}   +   (d-2)  \cot\theta  \frac{\partial \rho}{\partial\theta} \right].  
\label{eq:FPS}
\ee
Although the equation is one-dimensional, it depends on a dimension in which a system is embedded, in the 
above equation designated by $d$.   The other parameters in the above equation are $D_r$, representing 
the orientational diffusion constant, and $\theta$ designating the angle between ${\bf n}$ and the $x$-axis.

To simplify nomenclature, we introduce the units of time and length,  $\tau_k = 1/\mu K$ and $\lambda_k = v_0/\mu K$.  
This permits us to express the FP equation in dimensionless parameters:  
\be
0 =  \frac{\partial}{\partial z} \left[ \left( z  -   \cos\theta \right) \rho \right] 
+  \alpha  \left[ \frac{\partial^2 \rho}{\partial\theta^2}   +   (d-2)  \cot\theta  \frac{\partial \rho}{\partial\theta} \right],
\label{eq:FPS-Z}
\ee  
where 
$
\alpha = \frac{D_r}{\mu K} 
$
is the dimensionless  diffusion constant and $z = (\mu K /v_0) x$.    

A stationary distribution averaged over all orientations, the quantity of primary interest in this work, is defined as 
\be
p(z) = \int_0^{\pi} d\theta\, \rho( z ,\theta) \,  \sin^{d-2} \theta.  
\label{eq:pz-d}
\ee
In the absence of thermal fluctuations, particles are confined to the region $-1 \le z \le 1$, 
where $z=1$ implies $v_0  = \mu K x$, representing the balance between the swimming velocity and the 
harmonic force.

\section{recurrence relation}

As there is no simple method to solve the FP equation or to transform it into the differential equation for $p$, 
we proceed to transform it into the following recurrence relation:  
\ba
A_{l,m}   &=&   \left[ \frac{ \alpha m^2 - \alpha m   }{  l +  \alpha m^2 + (d-2)\alpha m  }  \right]  A_{l,m-2} \nonumber\\ 
&+& \left[ \frac{l}{  l +  \alpha m^2 + (d-2) \alpha m } \right]  A_{l-1,m+1}, 
\label{eq:Alm}
\ea
where 
\be
A_{l,m} = \langle z^l\cos^m\theta \rangle,  
\ee
and the angular brackets designate an averaging procedure 
$
\langle \dots \rangle = \int_{-1}^{1} dz \,  \int_0^{\pi} d\theta \sin^{d-2}\theta\, (\dots)\,  \rho(z,\theta).  
$

The recurrence relation in Eq. (\ref{eq:Alm}) was obtained by multiplying Eq. (\ref{eq:FPS-Z}) by 
$z^l\cos^m\theta$ followed by integration $\int_{-1}^{1} dz  \int_0^{\pi} d\theta \sin^{d-2}\theta$ and the application of the 
integration by parts.  The coefficients that are of special interest to us are $A_{2n,0}$ 
\be
A_{2n,0}  =  \int_{-1}^{1} dz\, z^{2n} p(z)  = \langle z^{2n} \rangle,
\label{eq:A2n0}
\ee
corresponding to even moments of $p$.

Initial conditions required to solve the recurrence relation are provided by the terms 
$A_{0,m} = \langle \cos^{m}\rangle =   \frac{\int_0^{\pi} d\theta \sin^{d-2}\theta\, \cos^m\theta}{\int_0^{\pi} d\theta \sin^{d-2}\theta }$, 
which after evaluation become 
\be
A_{0,m}  =
\begin{dcases*}
 \frac{\Gamma \left(\frac{d}{2}\right) \Gamma \left(\frac{m}{2}+\frac{1}{2}\right)}{\sqrt{\pi } 
 \Gamma \left(\frac{d}{2}+\frac{m}{2}\right)}, & \text{$m = $ even}\\
0, & \text{$m = $ odd} 
\end{dcases*}
\label{eq:A0m}
\ee
and which depend on a dimension $d$ in which the potential $u = \frac{1}{2} Kx^2$ is embedded.

The recurrence relation cannot be solved or reduced to a simpler form, and the terms $A_{l,m}$ are generated 
directly from Eq. (\ref{eq:Alm}).  Details how this is done are provided in Appendix (\ref{sec:app1}).  The moments 
generated from the recurrence relation and defined in Eq. (\ref{eq:A2n0}) are listed  in Table (\ref{table1}) for 
dimensions $d=2$ and $d=3$.  
\begin{table}[h!]
\centering
 \begin{tabular}{ l l l } 
  \hline
\multicolumn{2}{c}{$d=2$}\\
 \hline
 & \\[-1ex]
   ~~~  {$ \langle z^2\rangle = \frac{1}{2} \frac{1}{1+\alpha}$ }   \\ [1.ex] 
   ~~~  {$ \langle z^4\rangle = \frac{3}{8} \frac{3 + 4 \alpha}{ (1 + \alpha) (3 + \alpha ) (1 + 2\alpha)}$}    \\ [1.ex] 
   ~~~  { $ \langle z^6\rangle = \frac{5}{16} \frac{15 + 68 \alpha + 93 \alpha^2 + 36 \alpha^3} { (1 + \alpha)^2 (3 + \alpha) (5 + \alpha) (1 + 2 \alpha) (1 + 3 \alpha) } $}   \\ [1.ex] 
   ~~~  { $ \langle z^8\rangle = \frac{35}{128} \frac{ 315 + 2628 \alpha + 7873 \alpha^2 + 10312 \alpha^3 + 5856 \alpha^4 + 1152 \alpha^5 } 
   { (1 + \alpha)^2 (3 + \alpha) (5 + \alpha) (7 + \alpha) (1 + 2 \alpha) (3 + 2 \alpha) (1 + 3 \alpha) (1 +  4 \alpha) } $}   \\ [1.ex] 
\hline
\multicolumn{2}{c}{$d=3$}\\
 \hline
& \\[-1ex]
   ~~~  {$ \langle z^2\rangle = \frac{1}{3} \frac{1}{1+2\alpha}$ }   \\ [1.ex] 
   ~~~  {$ \langle z^4\rangle = \frac{1}{5} \frac{3 + 5 \alpha}{ (1 + 2\alpha) (3 + 2\alpha ) (1 + 3\alpha)}$}    \\ [1.ex] 
   ~~~  { $ \langle z^6\rangle = \frac{1}{7} \frac{30 + 167 \alpha + 287 \alpha^2 + 140 \alpha^3} { (1 + 2\alpha) (3 + 2\alpha) (5 + 2\alpha) (1 + 3 \alpha) (2 + 3 \alpha) (1+4\alpha) } $}   \\ [1.ex] 
   ~~~  { $ \langle z^8\rangle = \frac{1}{9} \frac{ 210 + 1859 \alpha + 5610 \alpha^2 + 6405 \alpha^3 + 2100 \alpha^4 } 
   { (1 + 2\alpha) (3 + 2\alpha) (5 + 2\alpha) (7 + 2\alpha) (1 + 3 \alpha) (2 + 3 \alpha) (1 + 4 \alpha) (1 +  5 \alpha) } $}   \\ [1.ex] 
\hline
\end{tabular}
\caption{Moments of a stationary distribution $p$ (defined in Eq. (\ref{eq:pz-d}))
generated by the recurrence relation in Eq. (\ref{eq:Alm}).  }
\label{table1}
\end{table}

A quick inspection of the table reveals that the respective moments of both tables become the same for $\alpha=2$ ($d=2$) 
and $\alpha=1/2$ ($d=3$), indicating that the distributions $p$ for those parameters are the same.  Interestingly, 
the remaining coefficients $A_{l,m}$  for this specific combination of $\alpha$'s are different, which implies that this
coincidence does not extend to other quantities or distributions.

\subsection{recovering distributions from moments}
\label{sec:rec-pz}


To recover $p(z)$ from the moments, we use the Fourier-Legendre series of $p$: 
\be
p(z) = \sum_{n=0}^{\infty} a_n P_{2n} (z), 
\label{eq:pz-Pn}
\ee
where $P_m$ are Legendre polynomials given by  
\be
P_m(z) = 2^m \sum_{k=0}^m z^k  {m \choose k}  {\frac{m+k-1}{2} \choose m}.  
\label{eq:P2n}
\ee
Note that the polynomials $P_m$ are defined on the interval $[-1,1]$, just as the distribution $p$.  
The coefficients of expansion $a_n$ are obtained from  
\be
a_n = \frac{ 4n + 1 }{2}  \int_{-1}^{1} dz\, P_{2n} (z) p(z).  
\ee
Together with Eq. (\ref{eq:P2n}) this yields 
\be
a_n = \frac{4 n+1}{2}  \left[ 2^{2 n} \sum _{k=0}^n \frac{  \langle z^{2 k} \rangle \,  \Gamma \left(k+n+\frac{1}{2}\right) } { (2 k)! (2 n-2 k)! \,  \Gamma \left(k-n+\frac{1}{2}\right)} \right], 
\label{eq:an}
\ee
which means that a given coefficient $a_n$ depends on $n-1$ initial moments.

Coefficients $a_n$ calcualted using Eq. (\ref{eq:an}) are listed in Table (\ref{table3}).  
\begin{table}[h!]
\centering
 \begin{tabular}{ l l l } 
  \hline
\multicolumn{2}{c}{$d=2$}\\
 \hline
 & \\[-1ex]
   ~~~  { $ a_0 = \frac{1}{2}$ }   \\ [1.ex] 
   ~~~  { $ a_1 =  \frac{5}{8} \frac{1-2\alpha}{ \alpha + 1} $}    \\ [1.ex] 
   ~~~  { $ a_2 =  \frac{27}{128}  \frac{9  - 60 \alpha - 8 \alpha^2 + 16 \alpha^3} { (\alpha + 1)(\alpha + 3) (2\alpha + 1) } $}   \\ [1.ex] 
   ~~~  { $ a_3 =  \frac{65}{512}  \frac{ 75 - 1010 \alpha + 855 \alpha^2 - 730 \alpha^3 - 1488 \alpha^4 + 32 \alpha^5 + 96 \alpha^6 }  { (\alpha + 1)^2 (\alpha + 3) (\alpha + 5)  (2\alpha + 1) (3\alpha + 1) } $}   \\ [1.ex] 
\hline
\multicolumn{2}{c}{$d=3$}\\
 \hline
& \\[-1ex]
   ~~~  { $ a_0 = \frac{1}{2}$ }   \\ [1.ex] 
   ~~~  { $ a_1 =  -\frac{5\alpha}{2} \frac{1}{2\alpha+1} $}    \\ [1.ex] 
   ~~~  { $ a_2 =  -\frac{27\alpha}{4}  \frac{  2 - 2 \alpha - 3 \alpha^2 } { (2\alpha + 1)(2\alpha + 3) (3\alpha + 1) } $}   \\ [1.ex] 
   ~~~  { $ a_3 =  -\frac{13 \alpha}{4} \frac{   60   -  146 \alpha  -  568 \alpha^2   -  230 \alpha^3   +  405 \alpha^4   +  180 \alpha^5  } {  (2 \alpha+1) (2 \alpha+3) (2 \alpha+5) (3 \alpha+1) (3 \alpha+2) (4 \alpha+1)} $}   \\ [1.ex] 
\hline
\end{tabular}
\caption{Coefficients $a_n$ of the Fourier-Legendre series in in Eq. (\ref{eq:pz-Pn}).  }
\label{table3}
\end{table}
We again bring attention to the fact that the corresponding coefficients in two tables become the same for 
$\alpha=2$ ($d=2$) and $\alpha=1/2$ ($d=3$).  This confirms the previously made assertion based on 
Table (\ref{table1}) that the distributions that the distributions for different dimensions become identical for 
this particular combination of parameters.  
In addition, we draw attention to the fact that for $\alpha = 0$ (for $d=3$) only the coefficient $a_0$ is non-zero.  
This implies that in this limit $p$ is uniform on the interval $[-1,1]$.  

{
To be useful, the Fourier-Legendre series in Eq. (9) needs to be truncated, 
$p =  \sum_{n=0}^{N_c} a_n P_{2n} (z)$.  As the truncation will introduce an error, we need to determine $N_c$
which is sufficient for an accurate recovery of distributions.  In Fig. (1) and Fig. (2) 
we plot a number of different distributions obtained from a truncated series
and compare it with $p$ obtained from numerical simulations based on the integration of a Langevin equation.  
The results show that the truncation at $N_c=20$ yields distributions that are indistinguishable from those 
obtained from simulations. 
}
\graphicspath{{figures/}}
\begin{figure}[hhhh] 
 \begin{center}
 \begin{tabular}{rrrr}
\hspace{-0.3cm}\includegraphics[height=0.15\textwidth,width=0.17\textwidth]{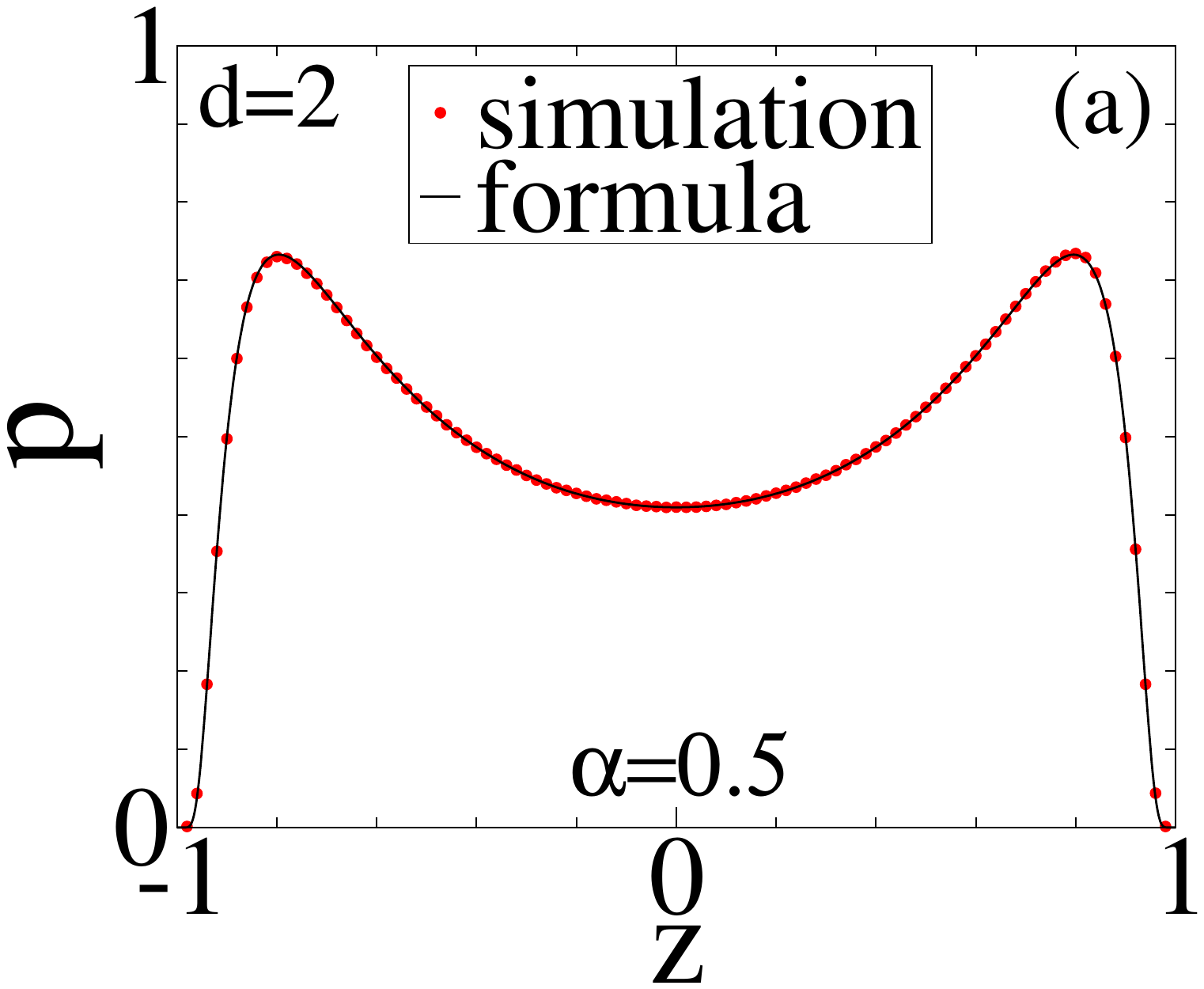} &
\hspace{-0.3cm}\includegraphics[height=0.15\textwidth,width=0.17\textwidth]{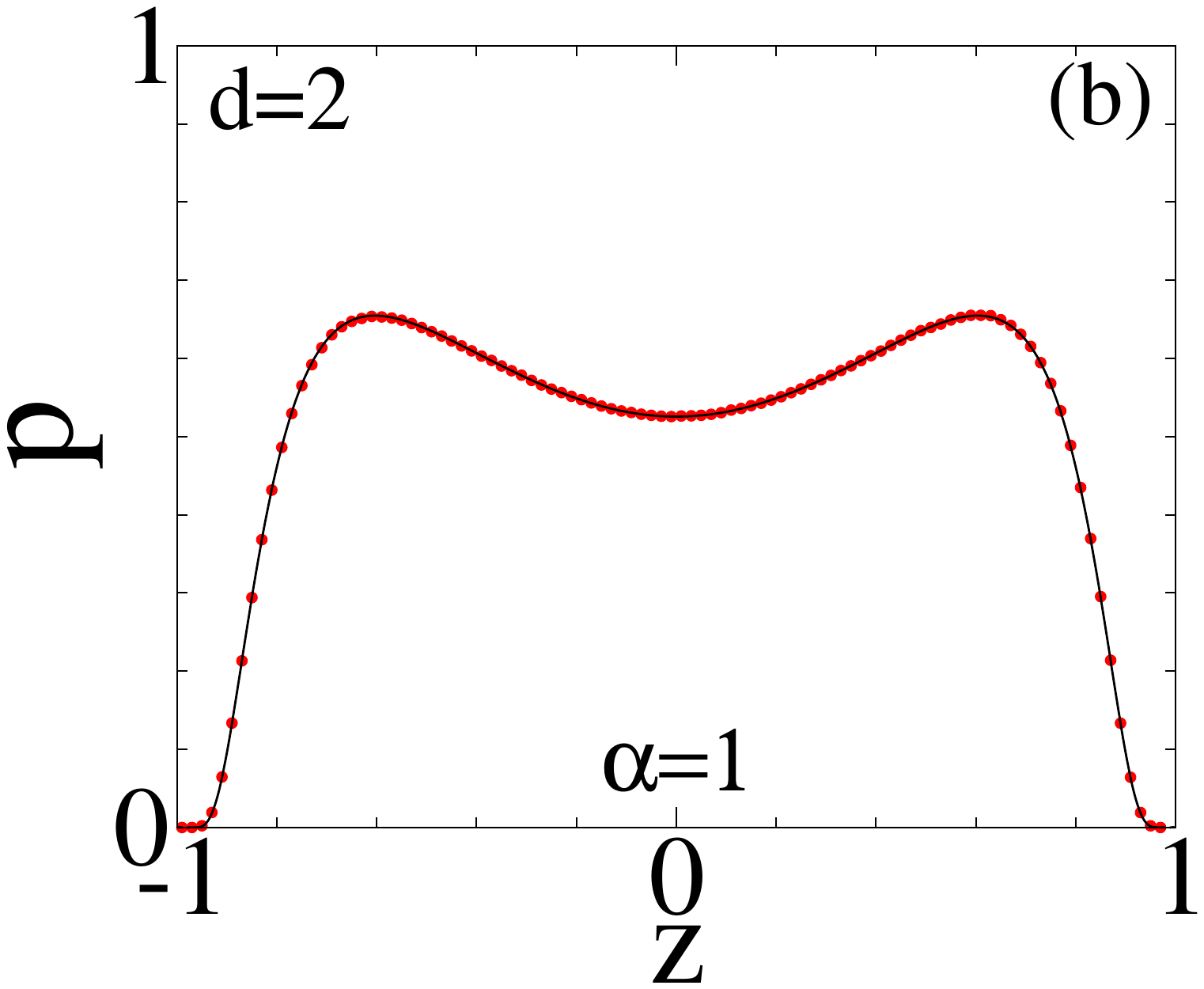} \\
\hspace{-0.2cm}\includegraphics[height=0.15\textwidth,width=0.17\textwidth]{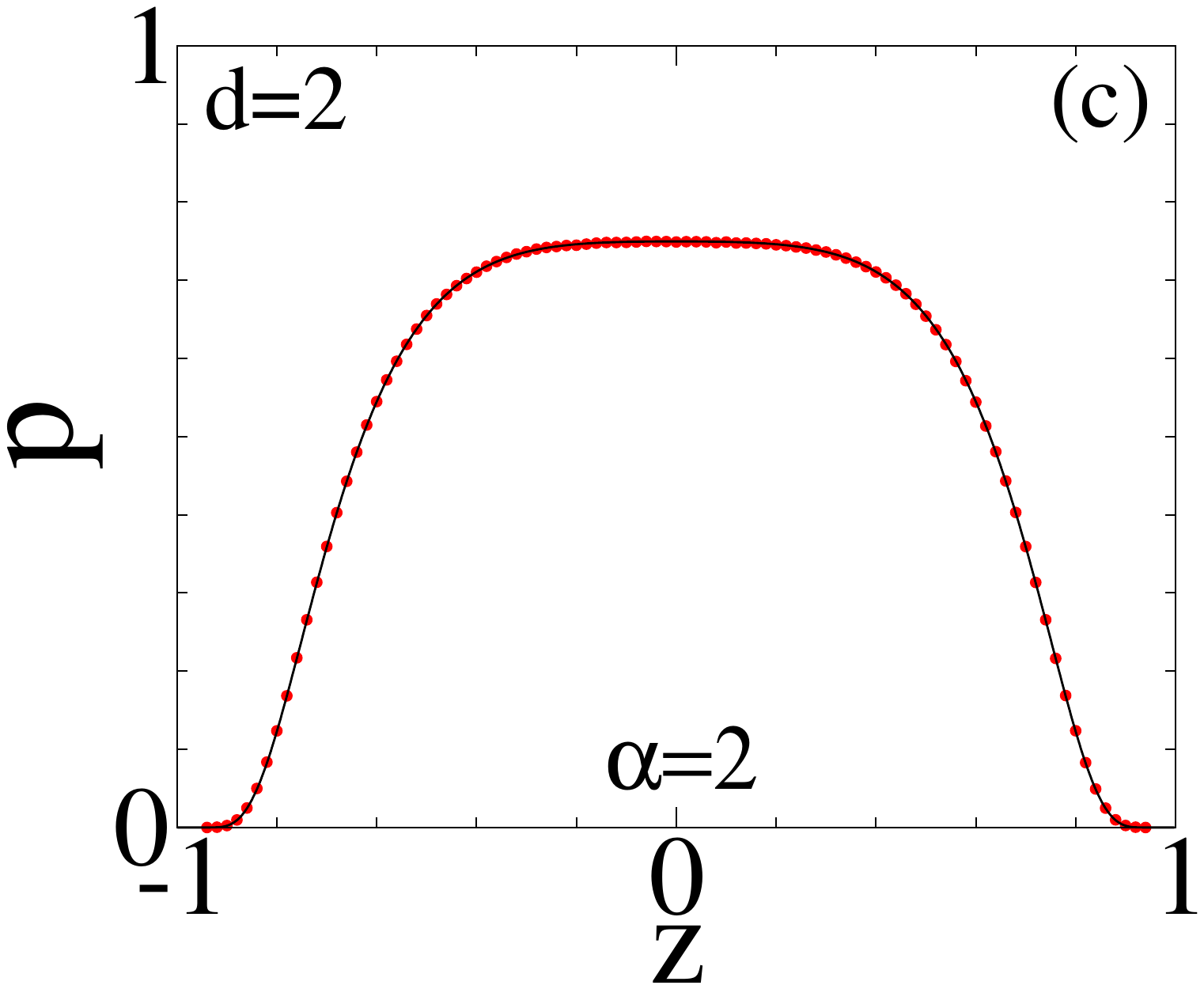} &
\hspace{-0.2cm}\includegraphics[height=0.15\textwidth,width=0.17\textwidth]{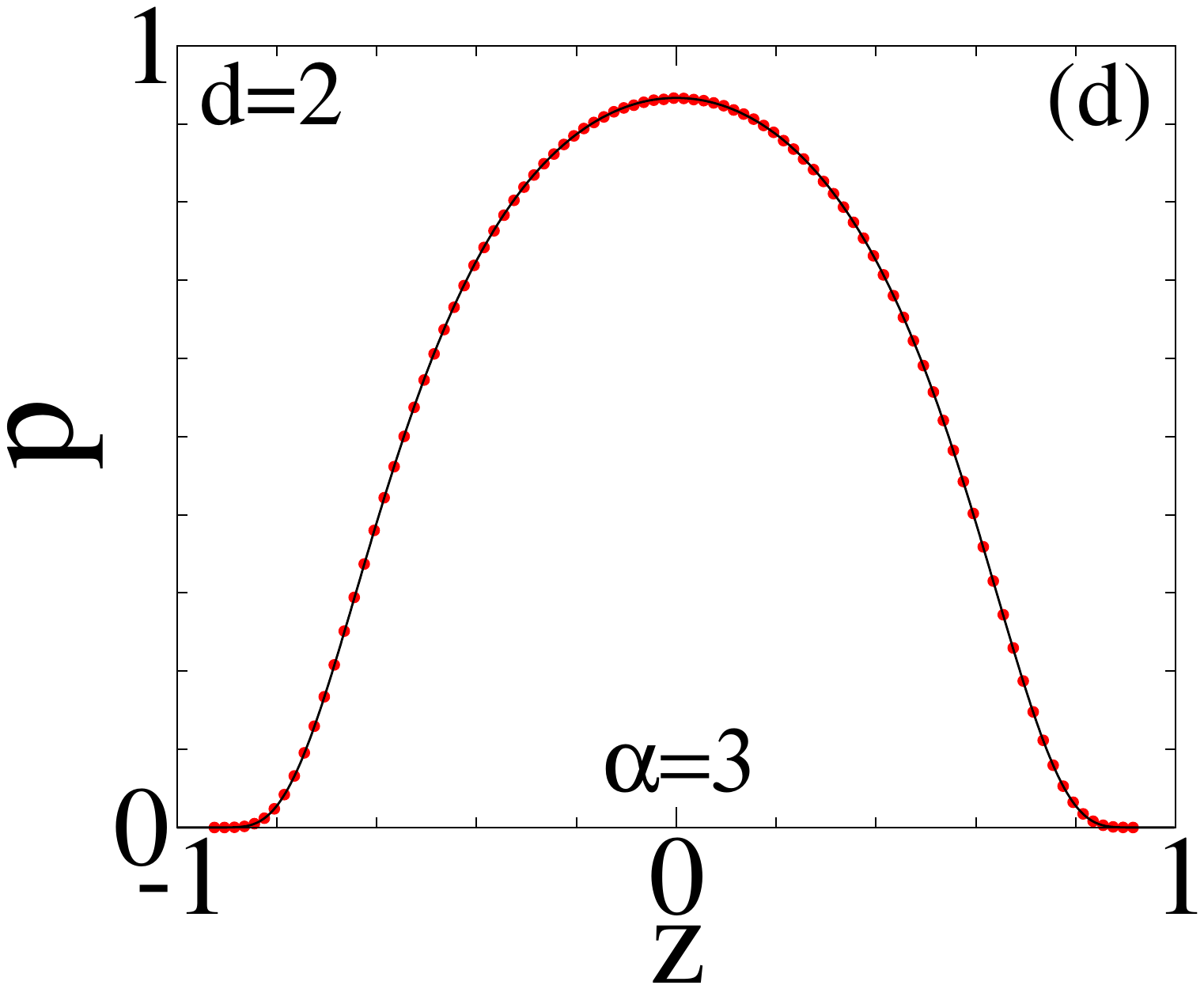} 
 \end{tabular}
 \end{center} 
\caption{Distributions $p$ calculated using the truncated Fourier-Legendre series $p =  \sum_{n=0}^{N_c} a_n P_{2n} (z)$, 
for $N_c=20$ initial terms.  The coefficients of expansion $a_n$ are defined in Eq. (\ref{eq:an}).  
The four distributions are (a) $\alpha=0.5$, (b) $\alpha=1$, (c) $\alpha=2$, (d) $\alpha=3$.}
\label{fig:pz2dl} 
\end{figure}
\graphicspath{{figures/}}
\begin{figure}[hhhh] 
 \begin{center}
 \begin{tabular}{rrrr}
\hspace{-0.3cm}\includegraphics[height=0.15\textwidth,width=0.17\textwidth]{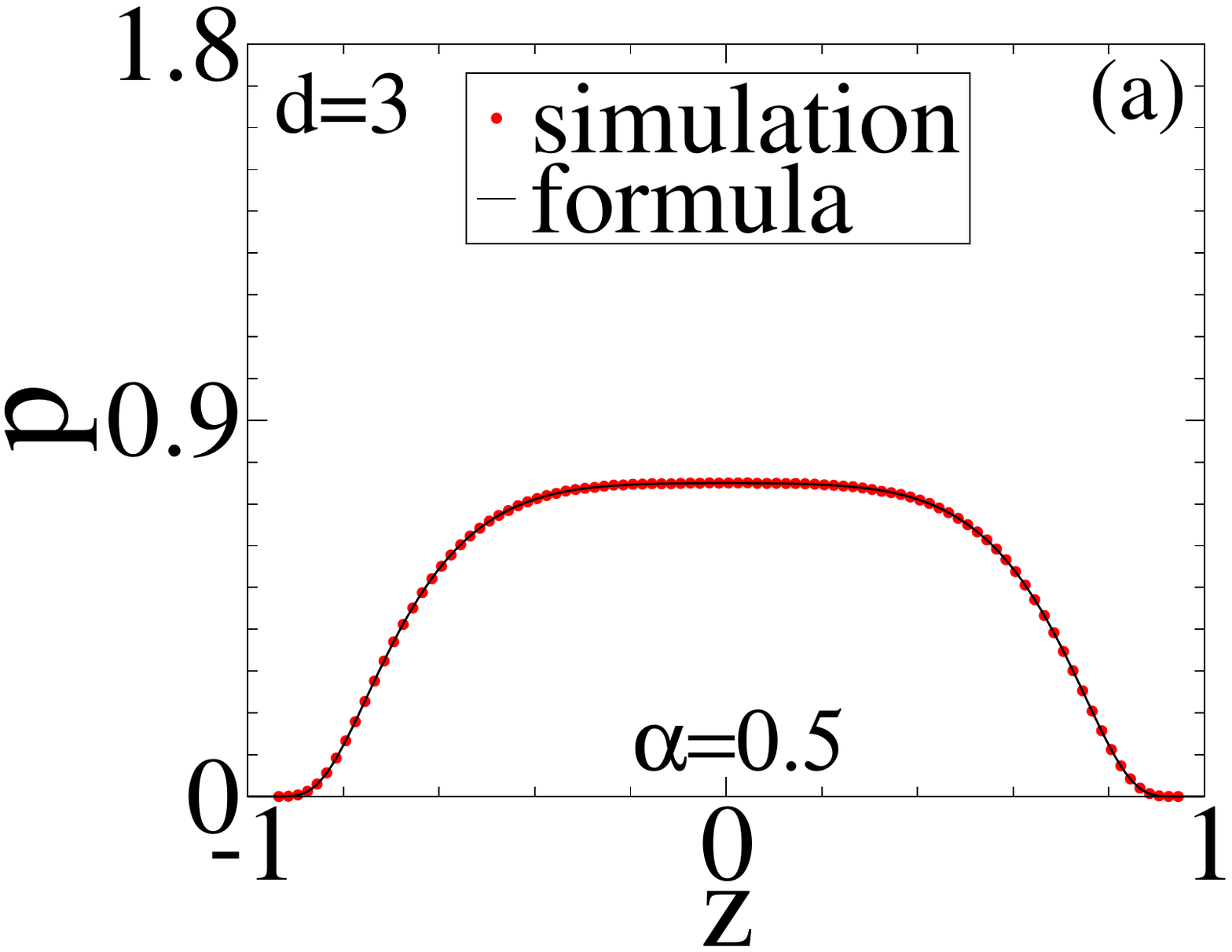} &
\hspace{-0.3cm}\includegraphics[height=0.15\textwidth,width=0.17\textwidth]{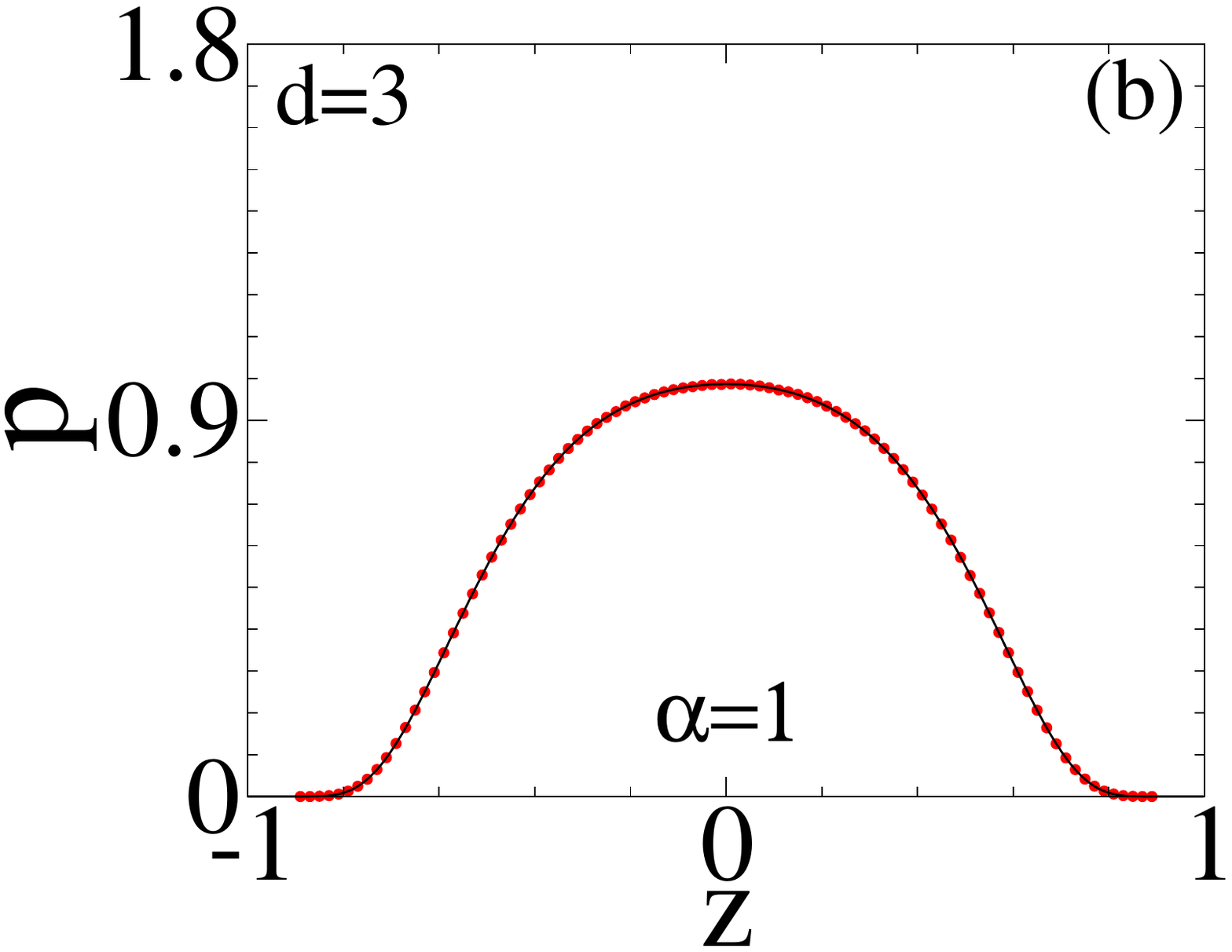} \\
\hspace{-0.2cm}\includegraphics[height=0.15\textwidth,width=0.17\textwidth]{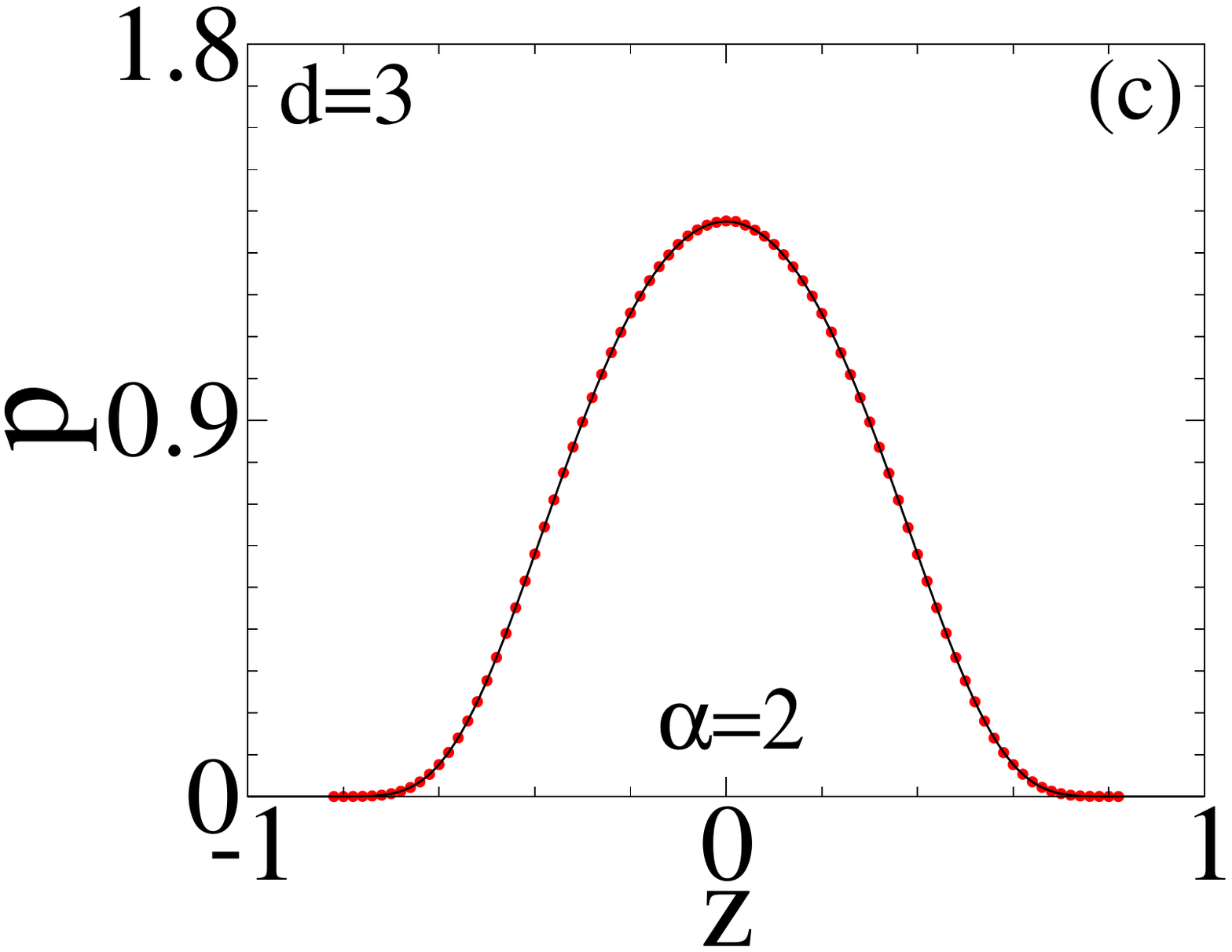} &
\hspace{-0.2cm}\includegraphics[height=0.15\textwidth,width=0.17\textwidth]{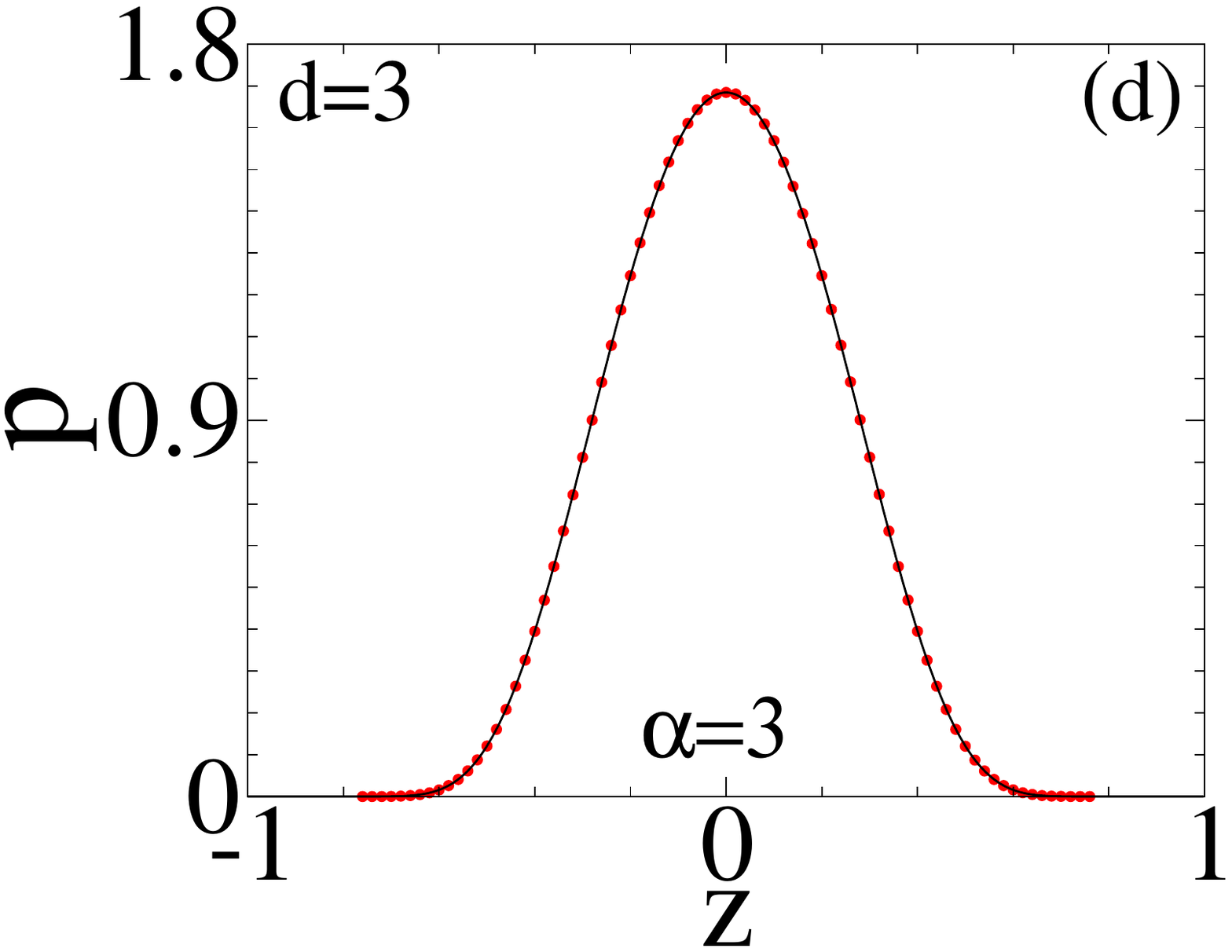} 
 \end{tabular}
 \end{center} 
\caption{Stationary istributions $p$ calculated using the truncated Fourier-Legendre series 
$p = \sum_{n=0}^{N_c} a_n P_{2n} (z)$, for $N_c=20$ initial terms.  The coefficients of expansion $a_n$ are 
defined in Eq. (\ref{eq:an}).  The four distributions are (a) $\alpha=0.5$, (b) $\alpha=1$, (c) $\alpha=2$, (d) $\alpha=3$.}
\label{fig:pz3dl} 
\end{figure}
{We bring special attention to distributions in Fig. (1) (c) and Fig. (2) (a).  These distributions 
are identical, although they are for different $d$ and $\alpha$.  This is in agreement with the moments listed 
in Table (1).  The corresponding moments for different $d$ in that table become identical for those two specific values of $\alpha$.   }

Computational advantage of the procedure for generating distributions from  the truncated Fourier-Legendre series 
over numerical simulations is that once equations are set up, the results are obtained instantaneously.  There is
an additional advantage of analytical tractability which permits us to identify certain properties of the distributions 
in exact manner, such as the fact that the distributions for $\alpha=2$ (for $d=2$) and $\alpha=1/2$ (for $d=3$) 
are identical.

While for $d=3$ distributions are convex for any $\alpha$, for $d=2$ we observe a crossover from  
a concave to convex shape at around $\alpha\approx 2$.  The precise point of crossover can be determined 
from the condition 
$$
\frac{d^2 p}{dz^2}\bigg|_{z=0} = 0.  
$$
Within the Taylor expansion $p = \sum_{n} c_n z^{2n}$, this corresponds to $c_1 = 0$, and 
within the Fourier-Legendre expansion, $c_1$ is given by 
\be
c_1  = \sum_{n=0}^{\infty}  a_n  \frac{  2^{2 n-1} \,  \Gamma \left(1+n+\frac{1}{2}\right) } { (2 n-2 )! \,  \Gamma \left(1-n+\frac{1}{2}\right)}. 
\label{eq:c2}
\ee

\begin{figure}[hhhh] 
 \begin{center}
 \begin{tabular}{rrrr}
\hspace{-0.3cm}\includegraphics[height=0.19\textwidth,width=0.23\textwidth]{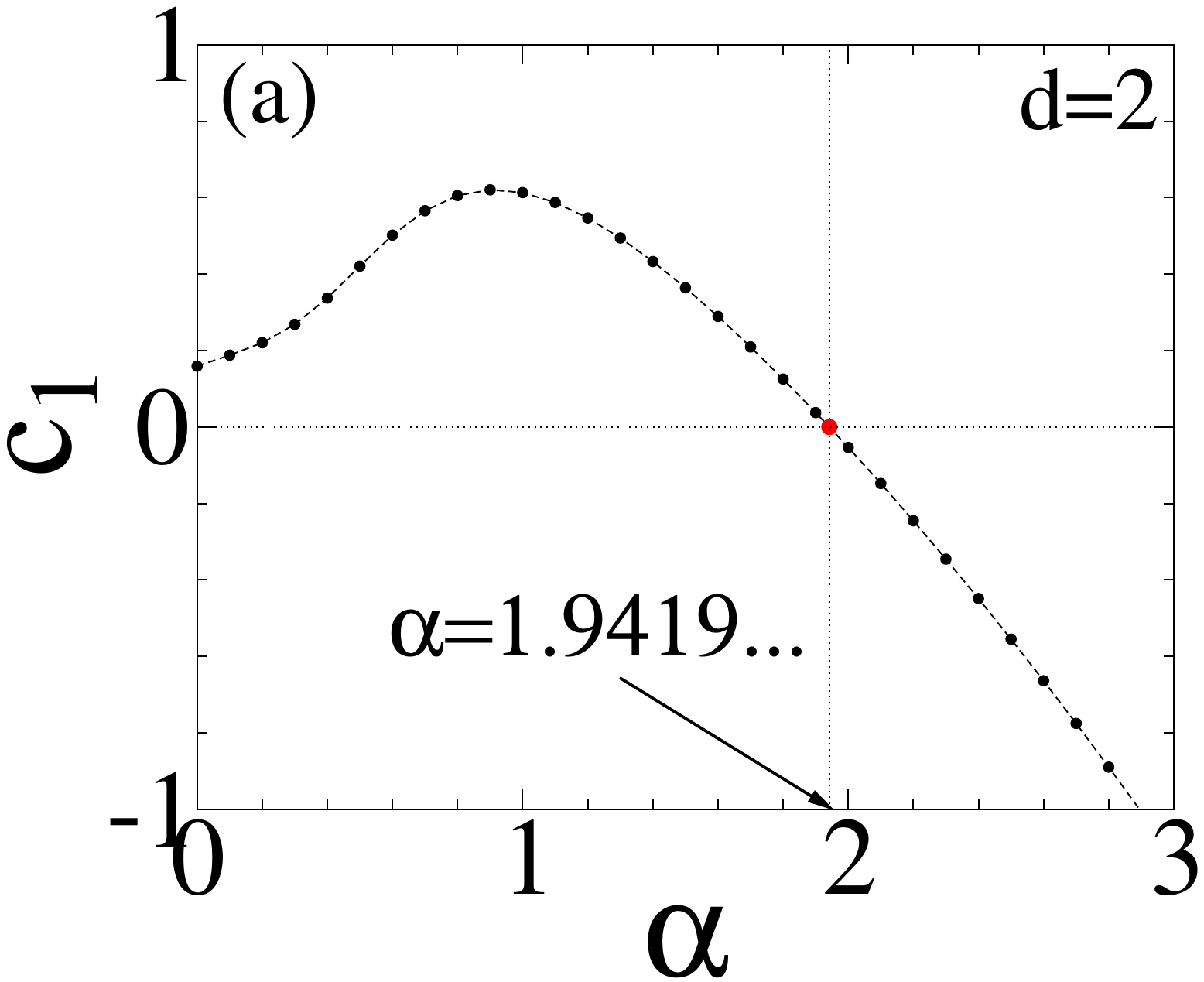} &
\hspace{0.cm}\includegraphics[height=0.19\textwidth,width=0.23\textwidth]{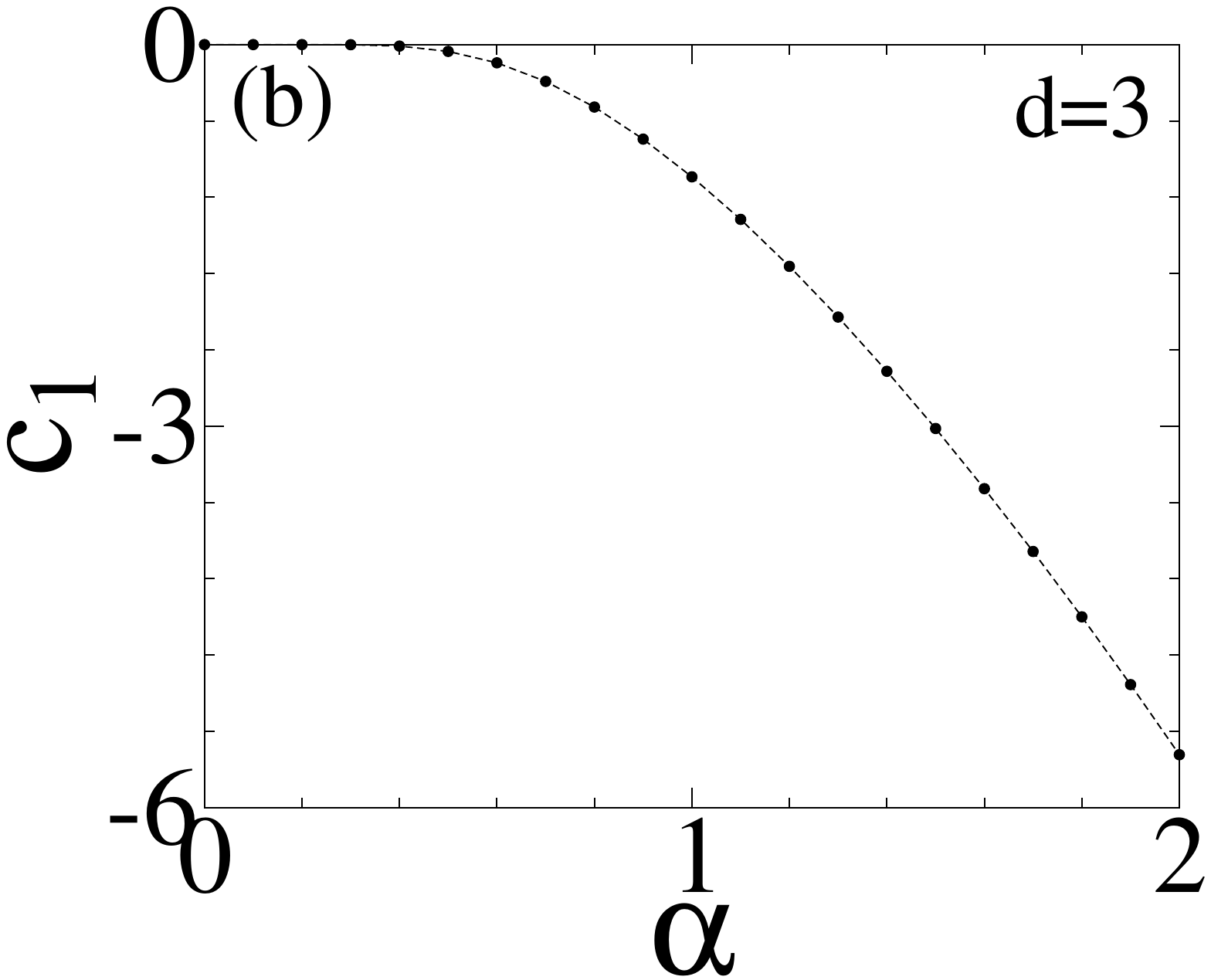} &
 \end{tabular}
 \end{center} 
\caption{The coefficient $c_1$ of the Taylor expansion $p = c_0 + c_1 z^2 + \dots$ as a function of $\alpha$ for (a) $d=2$
and (b) $d=3$.  The date points are obtained from Eq. (\ref{eq:c2}).  The dashed lines provide visual guidance between
these points.}
\label{fig:c2} 
\end{figure}

In Fig. (\ref{fig:c2}) we plot $c_1$ as a function of $\alpha$, calculated using Eq. (\ref{eq:c2}).  
{As in this case the aim is to obtain a precise value of $\alpha$, we use a larger number of terms of the 
series.  It was determined that to ensure the accuracy up to six significant digits, $N_c=100$ was sufficient. }
It was determined that for $d=2$, $c_1$ changes from positive to negative value at 
\be
\alpha = 1.94189\dots. 
\label{eq:alpha-c}
\ee
and $c_1$ of a system for $d=3$ is negative for $\alpha>0$ and monotonically 
decreasing with increasing $\alpha$, implying that $p$ is always convex.  
\graphicspath{{figures/}}

\subsection{distribution in $w$-space}

The interpretation of a shape of $p$ can be facilitated by looking at the corresponding distributions in a velocity 
space, where we define the velocity as $v = v_0\cos\theta -\mu Kx$, in dimensionless units given by 
\be
w = \cos\theta - z,
\label{eq:w}
\ee
where $w=v/v_0$.  Using Eq. (\ref{eq:w}), the moments $\langle w^{2n} \rangle$ can be related to the coefficients 
$A_{l,m} = \langle z^{l} \cos^m \theta\rangle$
\be
\langle w^{2n} \rangle = \sum_{k=0}^{2n} \frac{(-1)^k (2 n)!  }{k! (2 n-k)!} A_{2n-k,k}.  
\label{eq:w2n}
\ee

As in the case of $p$, the distribution $p_w$ are recovered using the Fourier-Legendre expansion:  
\be
p_w  =   \sum_{n=0}^{\infty} a_n P_{2n} \left( w/2 \right). 
\label{eq:pw-Pn}
\ee
Because both $z$ and $\cos\theta$ are defined on $[-1,1]$, $p_w$ is defined on $w\in[-2,2]$ as a result of Eq. (\ref{eq:w}).  
This requires that the Legendre polynomials be scaled as $P_{2n}(w/2)$.  The coefficients of expansion in this situation 
are given by 
\be
a_n = \frac{4 n+1}{2}  \left[ 2^{2 n} \sum _{k=0}^n \frac{  \langle w^{2 k} \rangle}{2^{2k+1}} \frac{  \Gamma \left(k+n+\frac{1}{2}\right) } { (2 k)! (2 n-2 k)! \,  \Gamma \left(k-n+\frac{1}{2}\right)} \right].  
\label{eq:an-pw}
\ee

\begin{figure}[hhhh] 
 \begin{center}
 \begin{tabular}{rrrr}
\hspace{-0.3cm}\includegraphics[height=0.15\textwidth,width=0.17\textwidth]{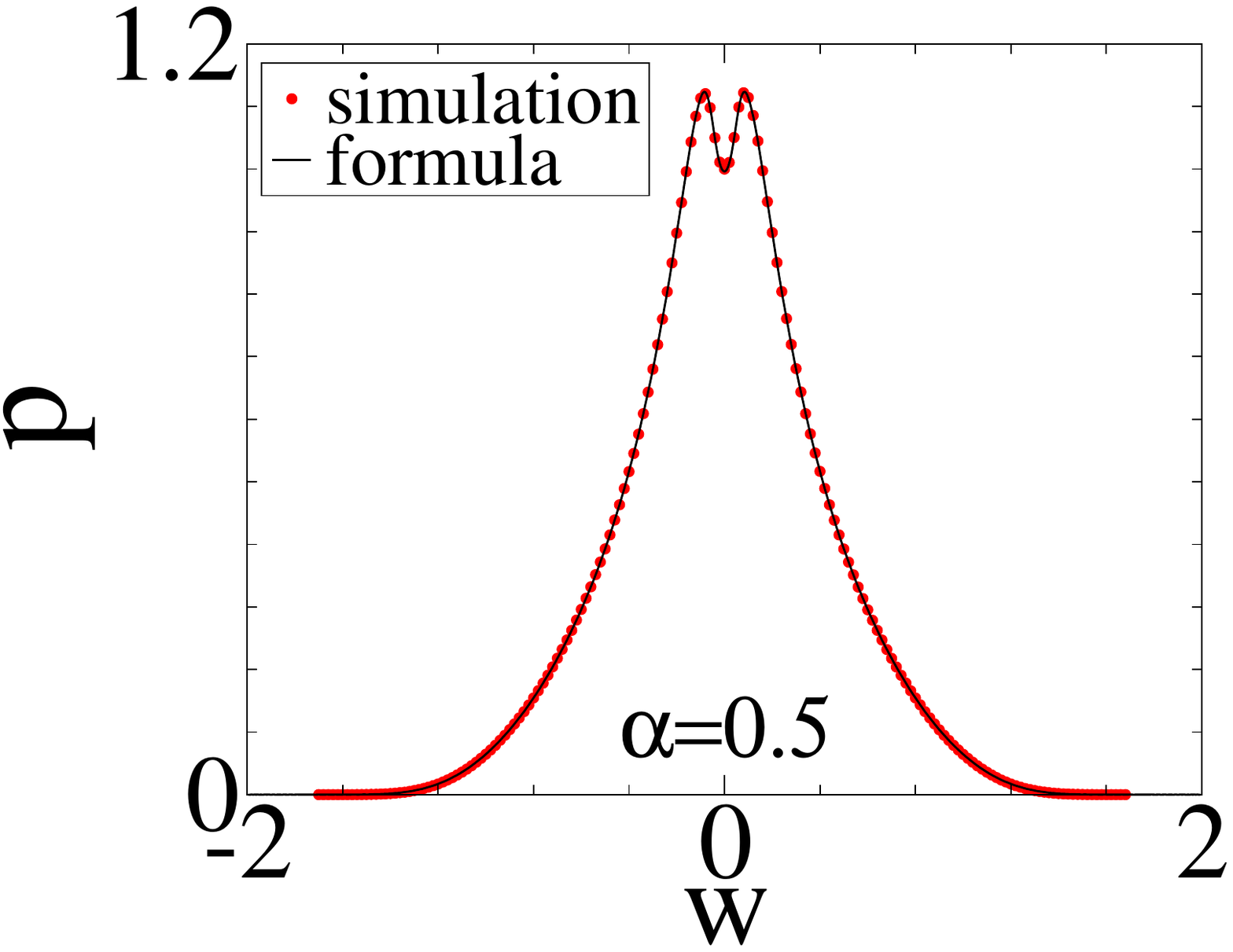} &
\hspace{-0.3cm}\includegraphics[height=0.15\textwidth,width=0.17\textwidth]{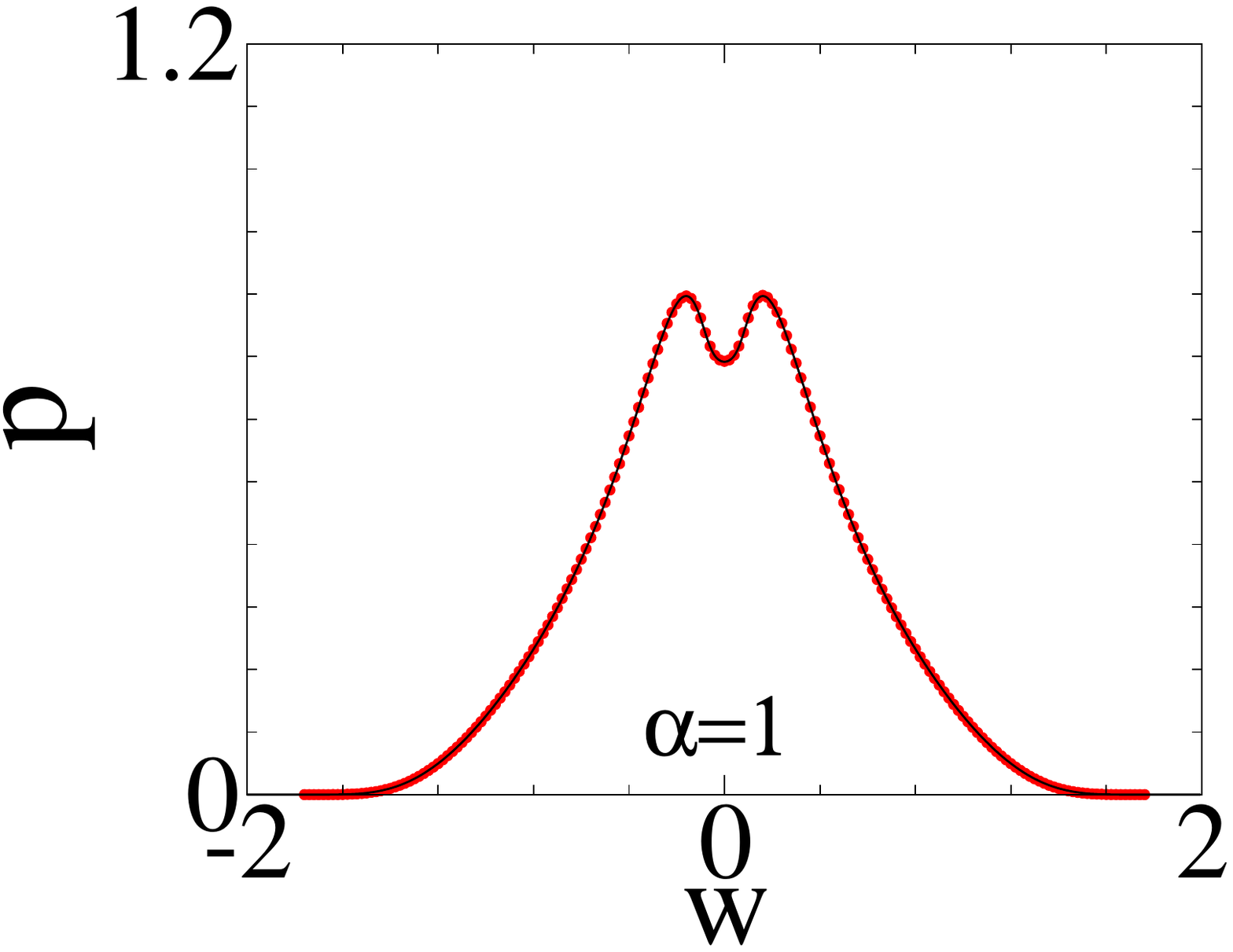} \\
\hspace{-0.2cm}\includegraphics[height=0.15\textwidth,width=0.17\textwidth]{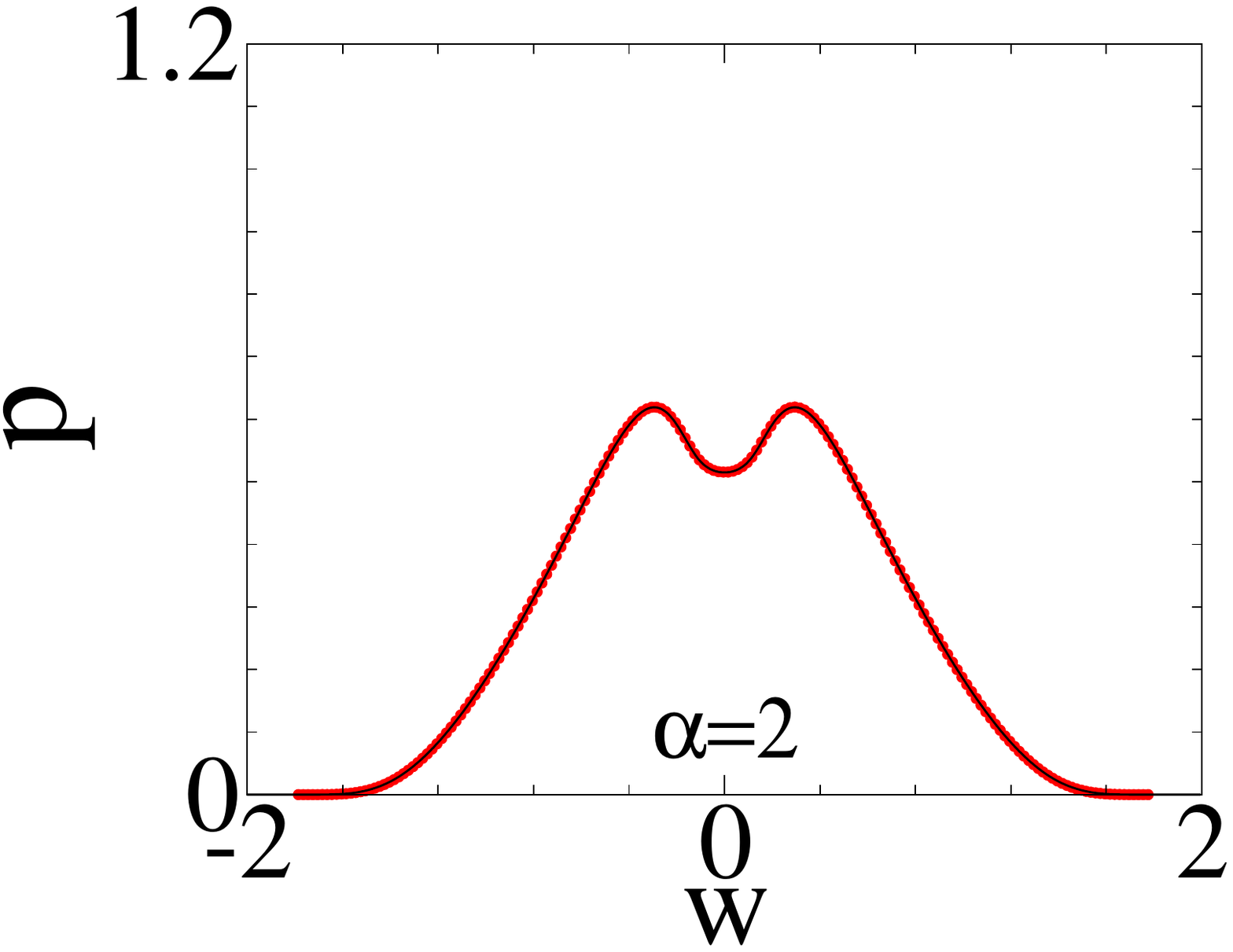} &
\hspace{-0.2cm}\includegraphics[height=0.15\textwidth,width=0.17\textwidth]{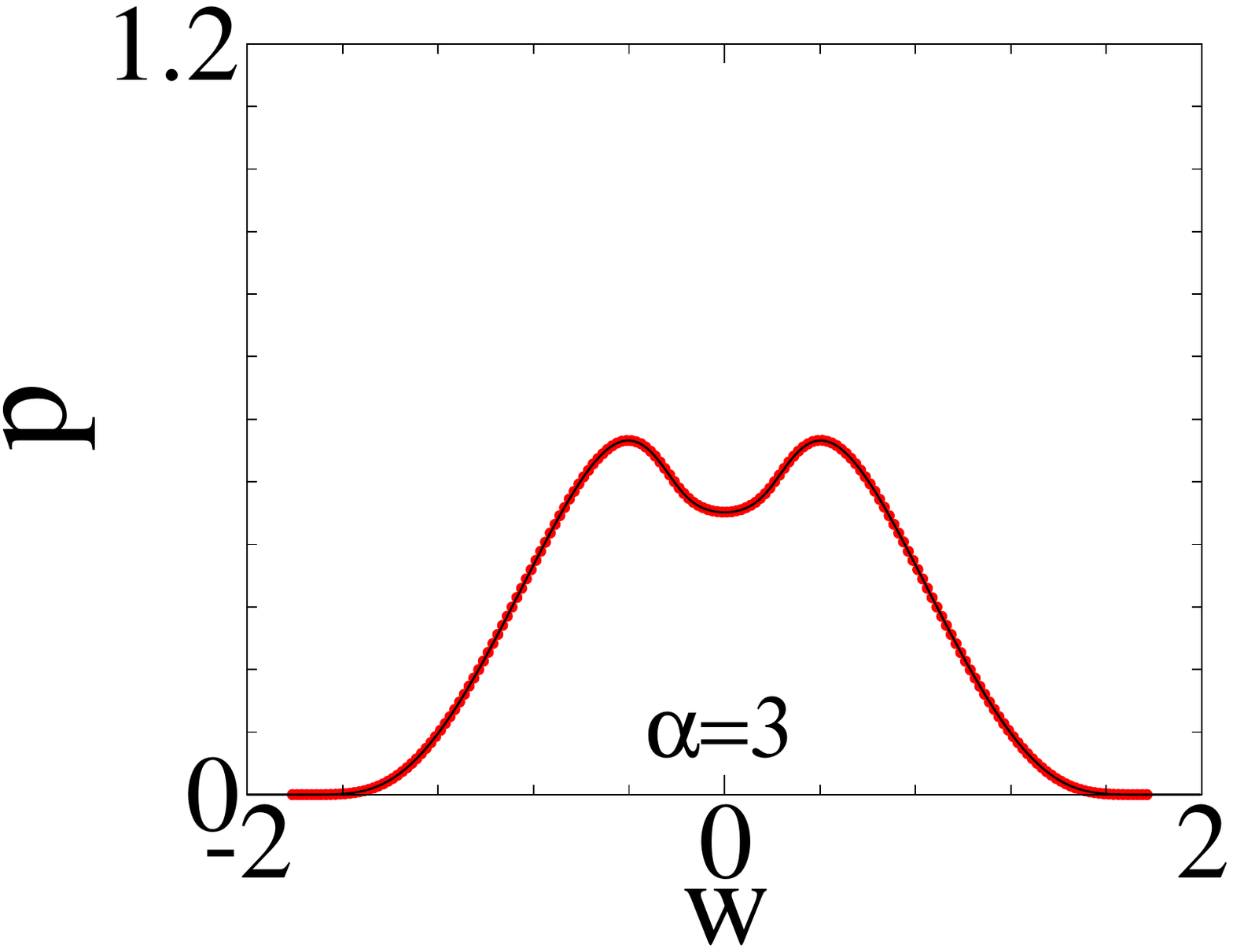} 
 \end{tabular}
 \end{center} 
\caption{Stationary distributions in the velocity space obtained from the truncated Fourier-Legendre series 
$p_w =   \sum_{n=0}^{N_c} a_n P_{2n} (w/2)$, for $N_c=20$ and a system dimension $d=2$.  }
\label{fig:pv2d} 
\end{figure}

\begin{figure}[hhhh] 
 \begin{center}
 \begin{tabular}{rrrr}
\hspace{-0.3cm}\includegraphics[height=0.15\textwidth,width=0.17\textwidth]{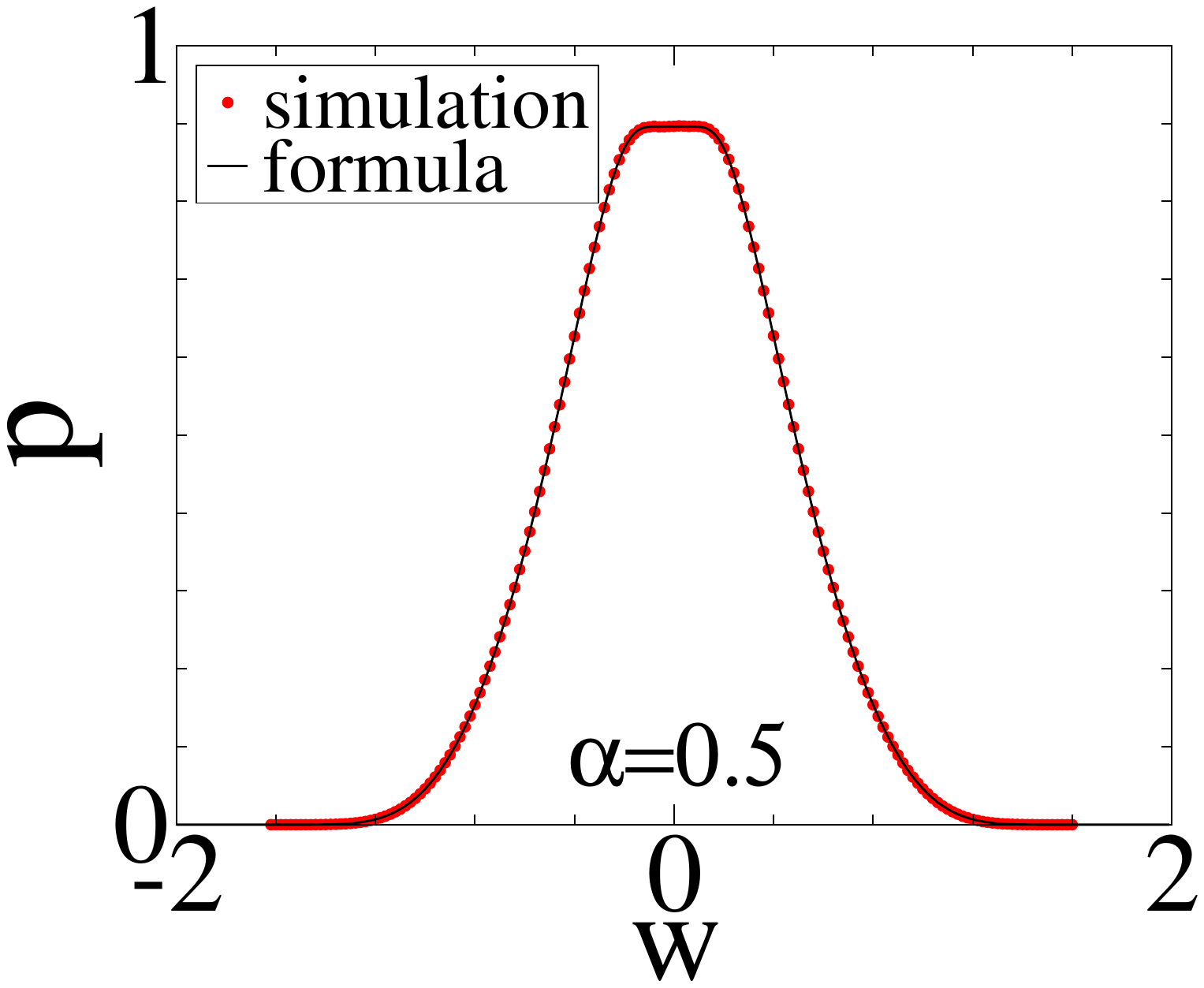} &
\hspace{-0.3cm}\includegraphics[height=0.15\textwidth,width=0.17\textwidth]{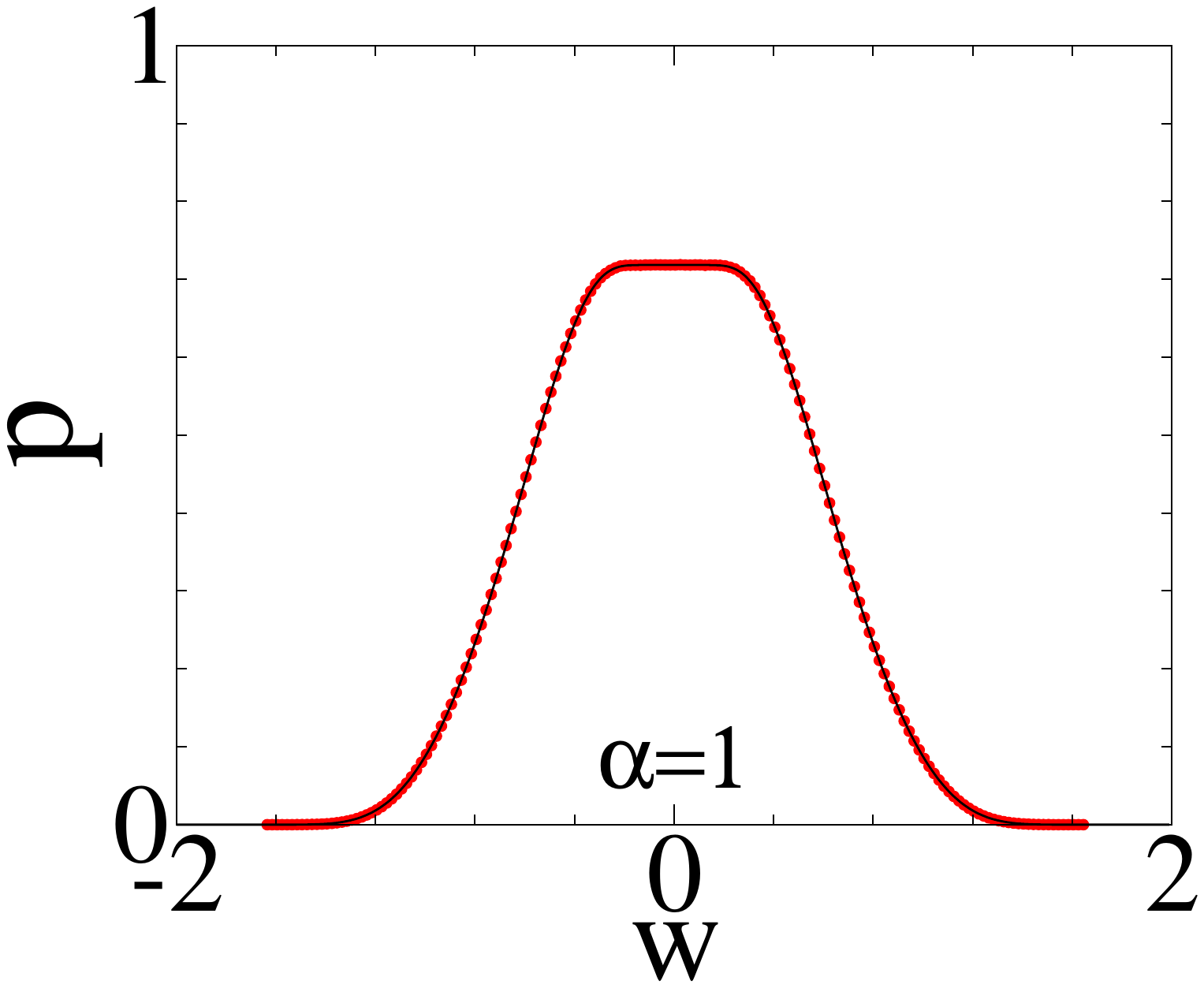} \\
\hspace{-0.2cm}\includegraphics[height=0.15\textwidth,width=0.17\textwidth]{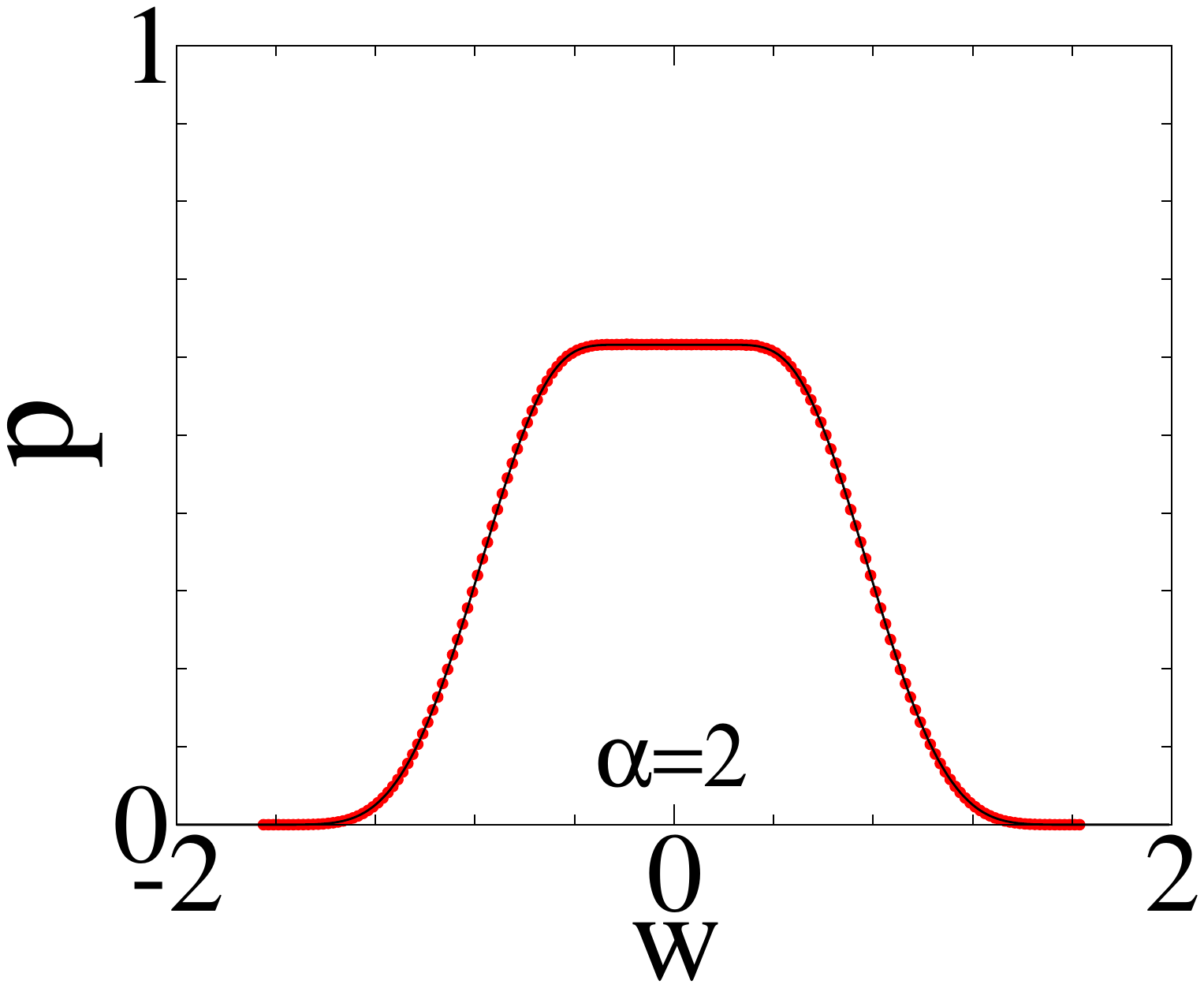} &
\hspace{-0.2cm}\includegraphics[height=0.15\textwidth,width=0.17\textwidth]{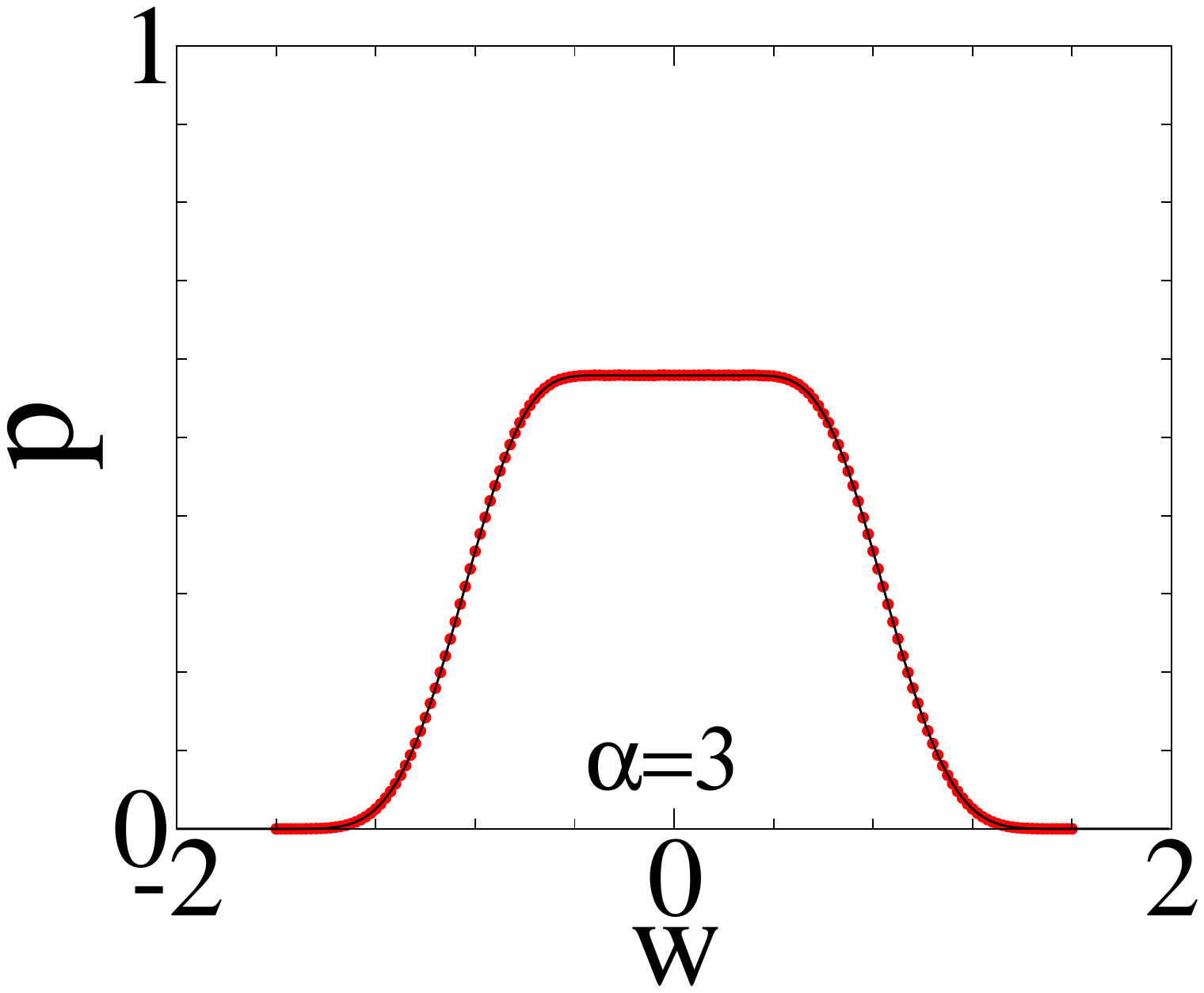} 
 \end{tabular}
 \end{center} 
\caption{Stationary distributions in the velocity space obtained from the truncated Fourier-Legendre series 
$p_w =   \sum_{n=0}^{N_c} a_n P_{2n} (w/2)$, for $N_c=20$ and a system dimension $d=3$. }
\label{fig:pv3d} 
\end{figure}

The expansion in Eq. (\ref{eq:pw-Pn}), together with the coefficients in Eq. (\ref{eq:an-pw}) and the moments
in Eq. (\ref{eq:w2n}), permits the recovery of $p_w$.  Fig. (\ref{fig:pv2d}) and Fig. (\ref{fig:pv3d}) shows some 
of those distributions for $d=2$ and $d=3$, respectively.  
\graphicspath{{figures/}}
\graphicspath{{figures/}}
 
There is a qualitative difference between distributions for different dimensions.  For $d=2$, distributions exhibit 
a scissor-like shape, and for $d=3$, distributions have a flat top.  
To understand different shapes of $p_w$ for different dimensions, it helps to consider the limit $\alpha\to \infty$.  
In this limit, the fast changing direction of a swimming velocity prevents a particle to move.  As a result, a particle 
becomes trapped at a potential minimum, $p\approx \delta(z)$, and any contributions of a potential force 
vanish so that $w \approx \cos\theta$.  Since orientations are uniformly distributed on $\theta\in[0,2\pi]$, in $d=2$, 
this leads to $\lim_{\alpha\to\infty} p_w =  \pi^{-1}\sqrt{1-w^2}$, 
and in $d=3$, 
$\lim_{\alpha\to\infty} p_w = \frac{1}{2}$.  Both distributions in this limit are defined on $w\in [-1,1]$.  Observe that distributions in 
Fig. (\ref{fig:pv2d}) and Fig. (\ref{fig:pv3d}) tend to those limiting distributions as $\alpha$ increases.

\section{Isotropic harmonic potential}

So far all the analysis has been carried out for a linear harmonic potential $u=\frac{1}{2} Kx^2$.  
It is possible to extend those results to an isotropic potential $u = \frac{1}{2} Kr^2$.   We recall 
that for an isotropic potential the moments are defined as $\langle s^{2n} \rangle = 2\pi \int_0^{1} ds\, s p(s)$ 
and $\langle s^{2n} \rangle = 4\pi \int_0^{1} ds\, s^2 p(s)$, for $d=2$ and $d=3$, respectively, where 
$s = (\mu K/v_0)r$ is the dimensionless radial distance.  

It turns out that the moments $\langle z^{2n} \rangle$ and $\langle s^{2n} \rangle$ are related via the following 
formula: 
\be
\langle  s^{2n}  \rangle   =  
\frac{\Gamma \left(\frac{1}{2}\right) \Gamma \left(\frac{d}{2}+n\right)}{\Gamma \left(\frac{d}{2}\right) \Gamma \left(n+\frac{1}{2}\right)} \langle  z^{2n}  \rangle.
\label{eq:r2n}
\ee
This relation has been first determined for the RTP particles \cite{Frydel23b}, and in this work we show
that it also holds for ABP particles.  A more detailed proof of this relation is relegated to Appendix (\ref{sec:app3}).   
For $d=3$, the relation in Eq. (\ref{eq:r2n}) simplifies to 
$
\langle  s^{2n}  \rangle   =   (2n+1) \langle  z^{2n}  \rangle.  
$

\subsection{recovering $p(s)$}

As done previously in Sec. (\ref{sec:rec-pz}), we can use the moments $\langle s^{2n}\rangle$ in Eq. (\ref{eq:r2n})  
to recover the corresponding distribution $p(s)$.  We start with $d=3$, where we expand the quantity 
$4\pi s^2 p(s)$, since this is a 
normalized quantity: 
\be
4\pi s^2 p(s) = \sum_{n=0}^{\infty} a_n P_{2n} (s), 
\label{eq:pz-Pn-3d}
\ee
with coefficients of expansion given by
\be
a_n = 
(4n + 1)
 \left[ 2^{2 n} \sum _{k=0}^n \frac{ \langle s^{2 k} \rangle \,  \Gamma \left(k+n+\frac{1}{2}\right) } { (2 k)! (2 n-2 k)! \,  \Gamma \left(k-n+\frac{1}{2}\right)} \right].  
\ee

For $d=2$, we cannot use the same procedure since $2\pi s p(s)$ is an odd function on $[-1,1]$. 
The Fourier-Legendre expansion in this case is (see Appendix (\ref{sec:app4}) for more details)
\be
p(s) = \frac{1}{\pi} \sum_{n=0}^{\infty} a_n P_n(2s^2-1), 
\label{eq:pz-Pn-2d}
\ee
with the coefficients of expansion given by 
\be
a_n = 2^n (2n + 1)  \sum_{k=0}^n  {n \choose k}  {\frac{n+k-1}{2} \choose n}  \sum_{p=0}^k { k \choose p } 2^{p}  (-1)^{k-p}  \langle s^{2p} \rangle.
\ee

\begin{figure}[hhhh] 
 \begin{center}
 \begin{tabular}{rrrr}
\hspace{-0.3cm}\includegraphics[height=0.15\textwidth,width=0.17\textwidth]{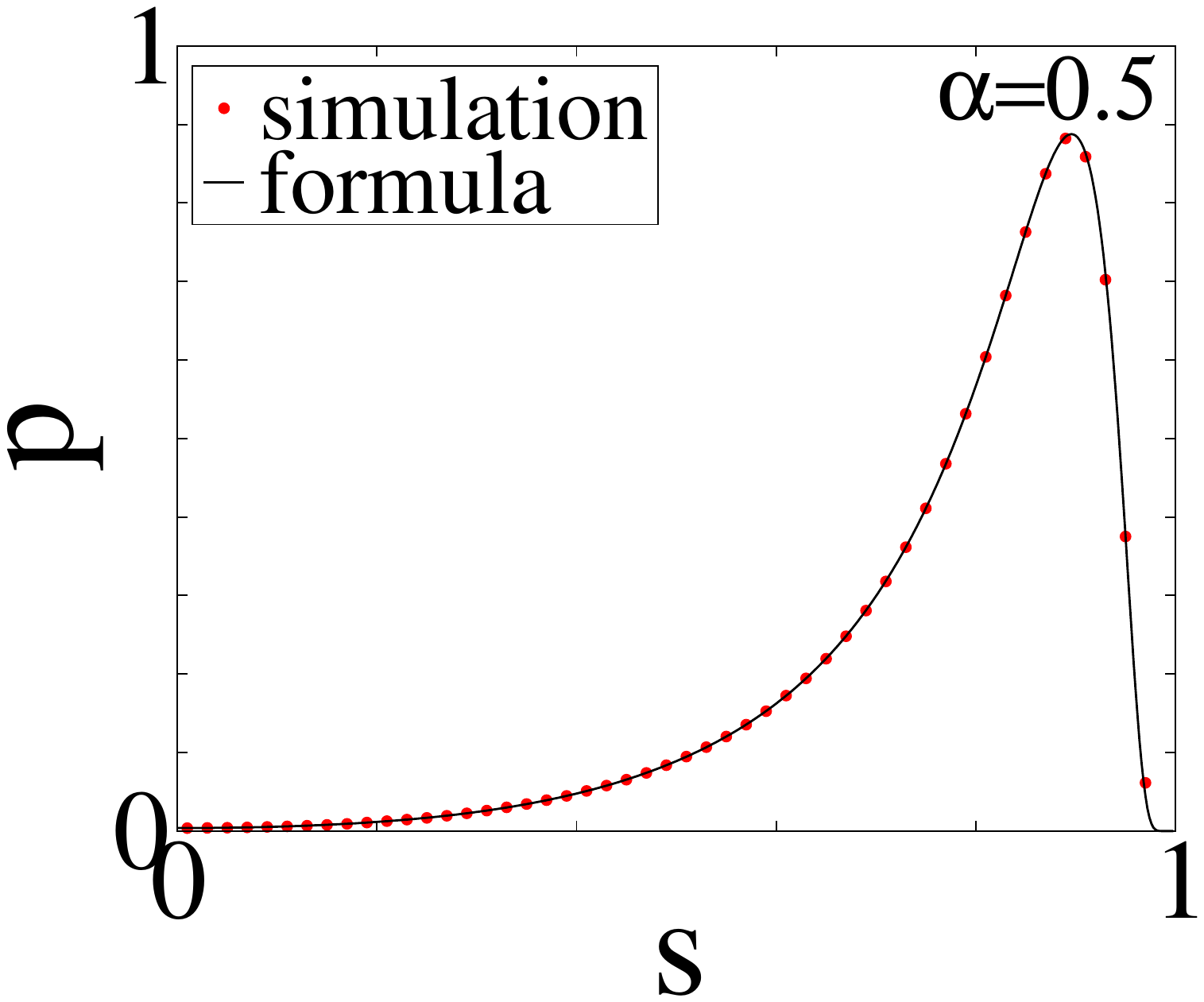} &
\hspace{-0.3cm}\includegraphics[height=0.15\textwidth,width=0.17\textwidth]{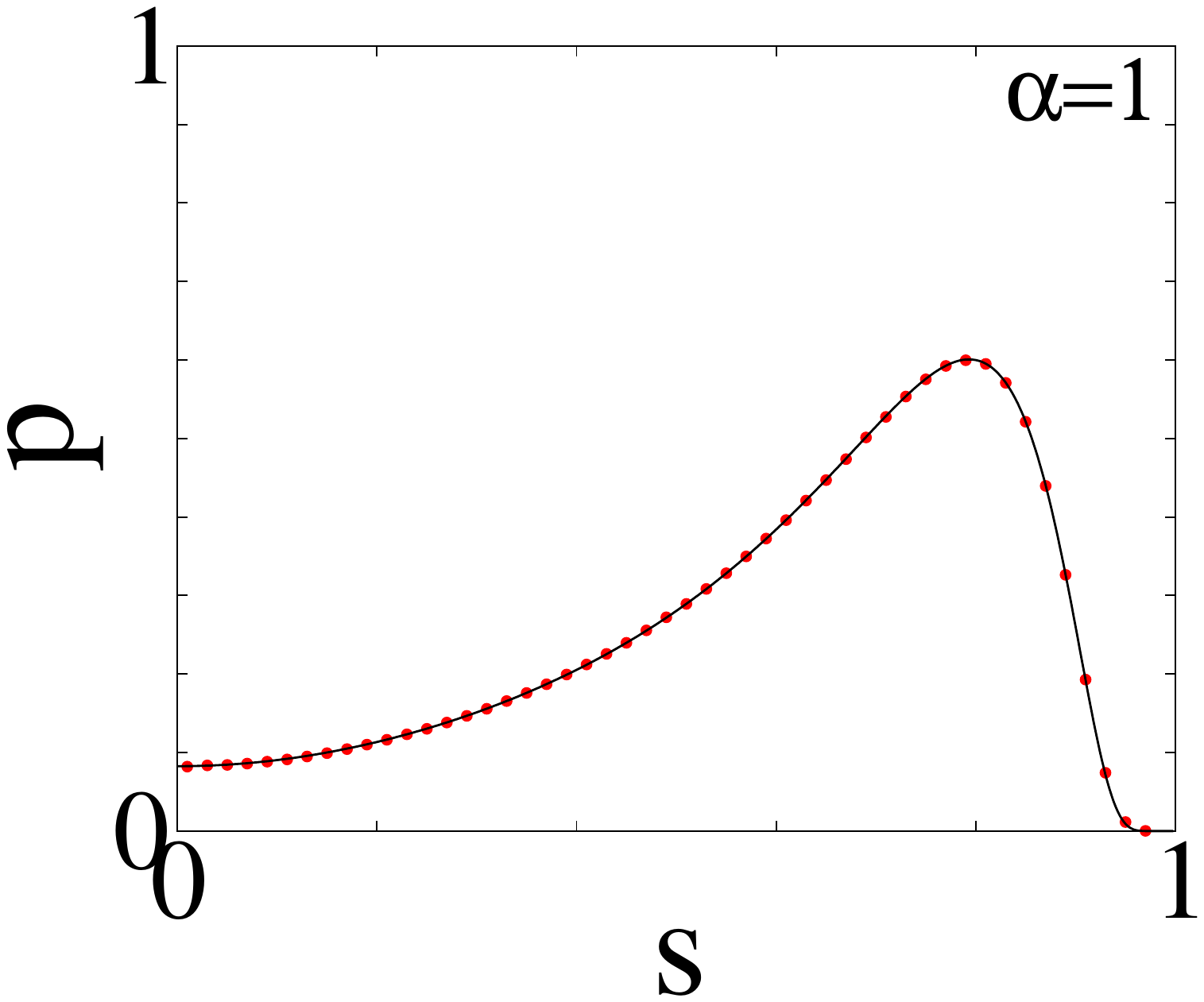} \\
\hspace{-0.2cm}\includegraphics[height=0.15\textwidth,width=0.17\textwidth]{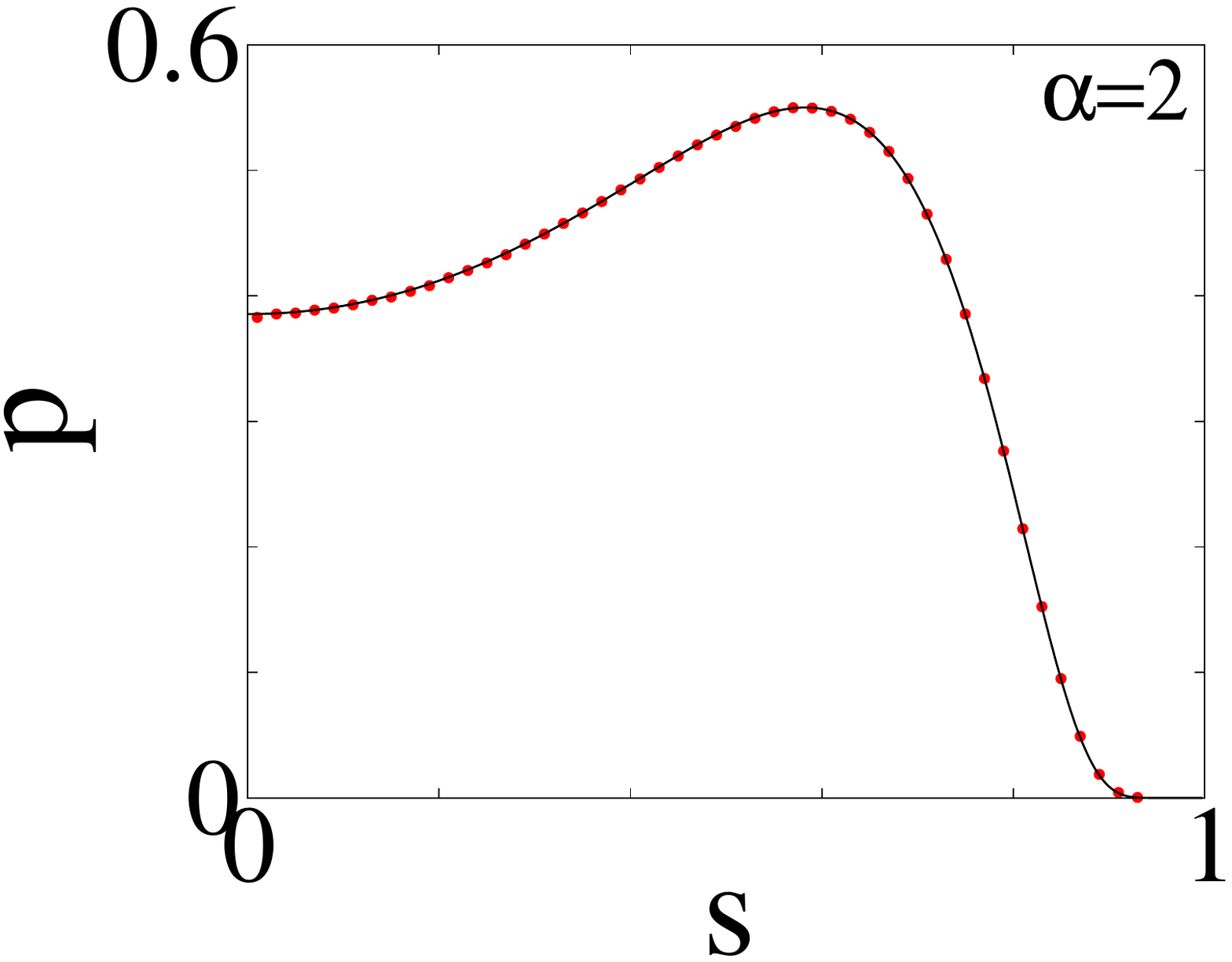} &
\hspace{-0.2cm}\includegraphics[height=0.15\textwidth,width=0.17\textwidth]{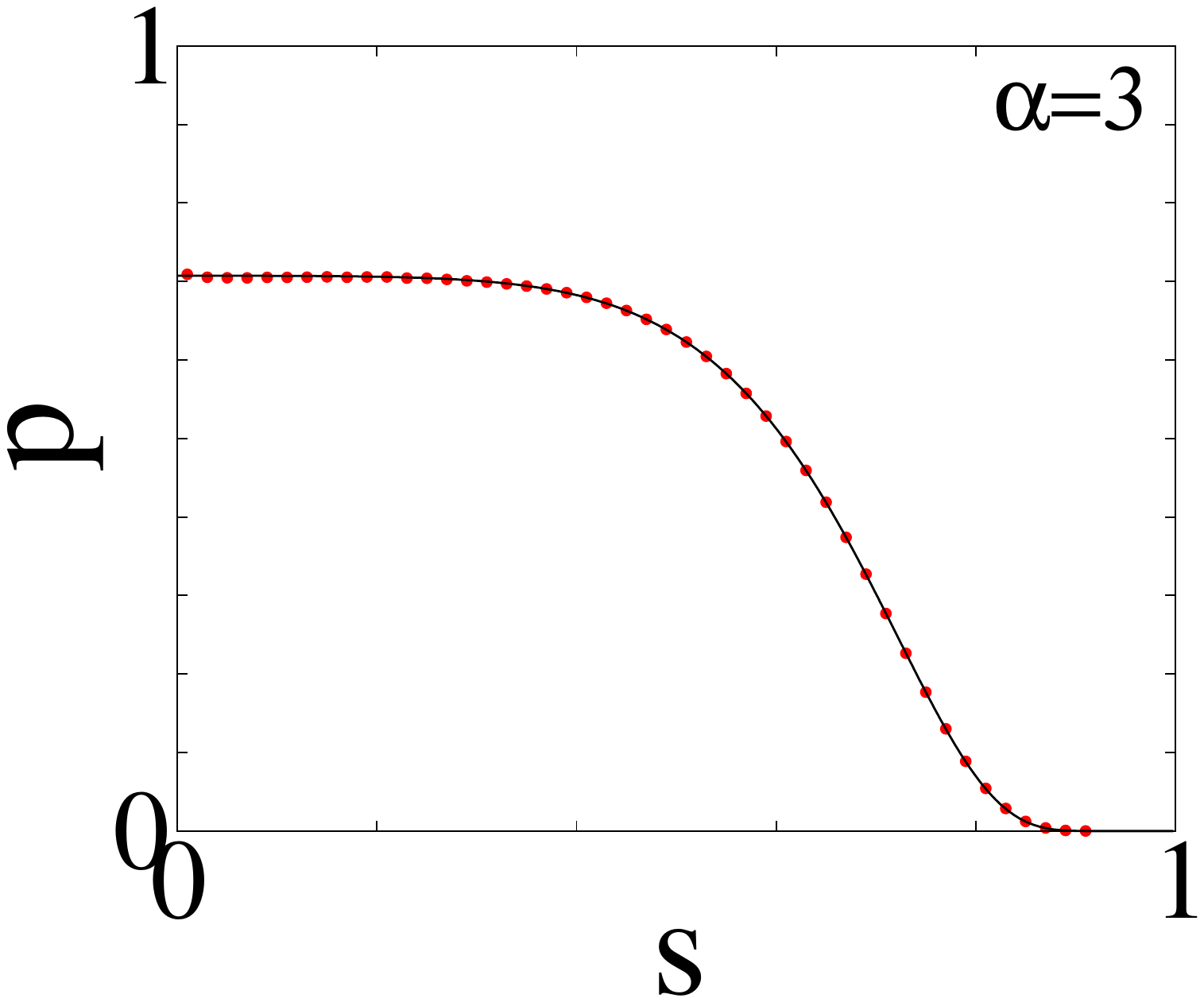} 
 \end{tabular}
 \end{center} 
\caption{Distributions $p(s)$ for $d=2$, calculated using the truncated Fourier-Legendre series in Eq. (\ref{eq:pz-Pn-2d}) for 
$N_c=20$.  }
\label{fig:pz2d} 
\end{figure}

\begin{figure}[hhhh] 
 \begin{center}
 \begin{tabular}{rrrr}
\hspace{-0.3cm}\includegraphics[height=0.15\textwidth,width=0.17\textwidth]{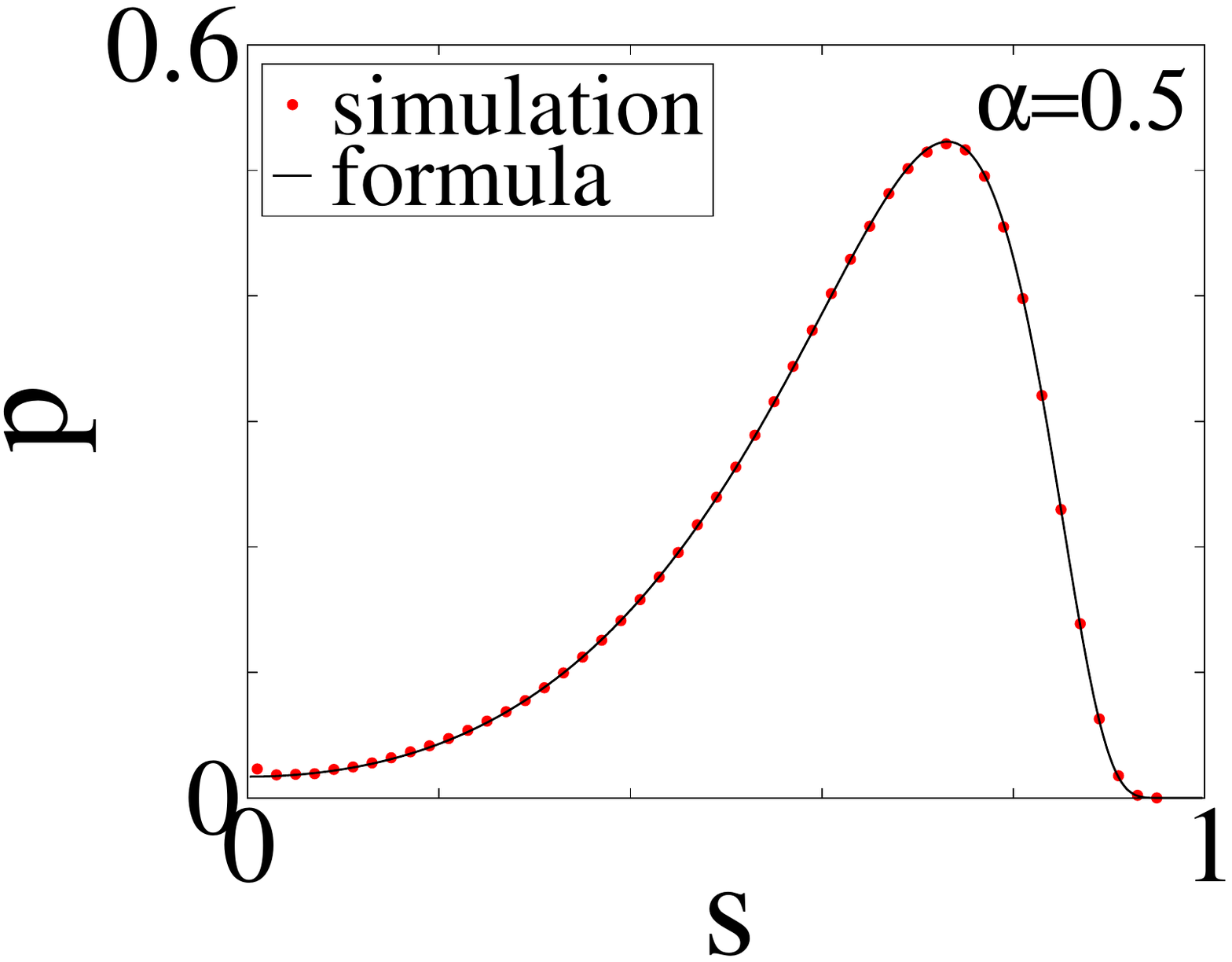} &
\hspace{-0.3cm}\includegraphics[height=0.15\textwidth,width=0.17\textwidth]{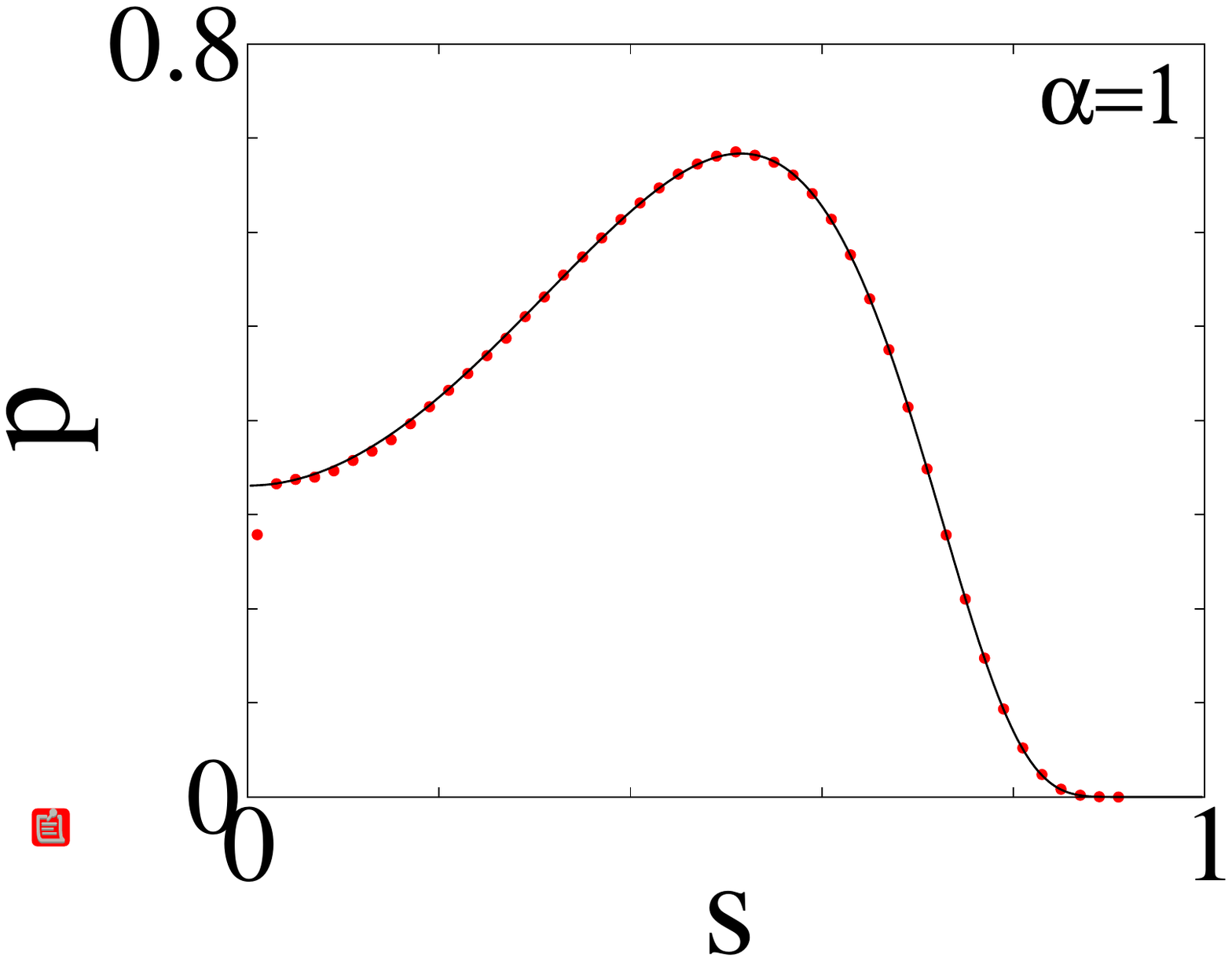} \\
\hspace{-0.2cm}\includegraphics[height=0.15\textwidth,width=0.17\textwidth]{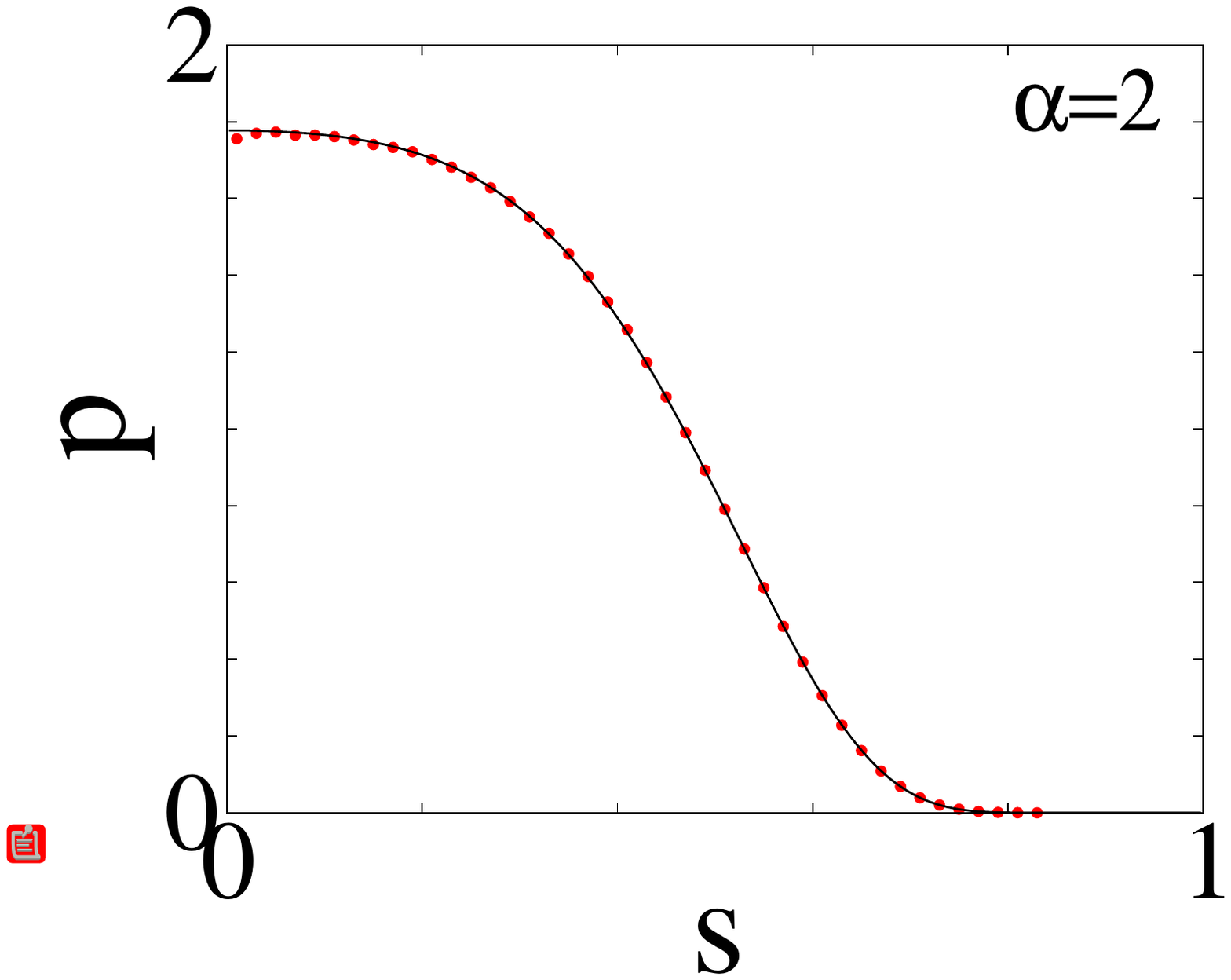} &
\hspace{-0.2cm}\includegraphics[height=0.15\textwidth,width=0.17\textwidth]{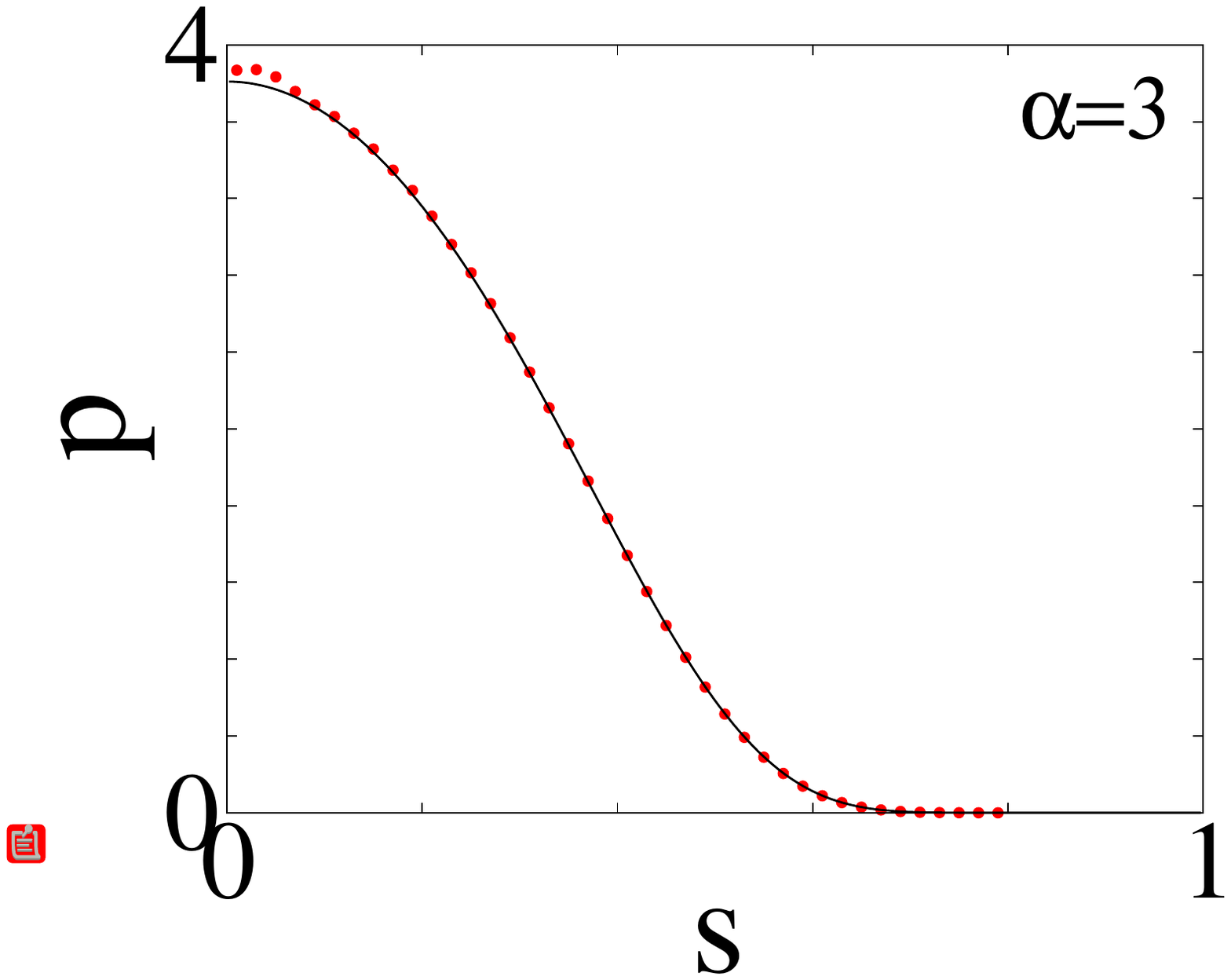} 
 \end{tabular}
 \end{center} 
\caption{Distributions $p(s)$ for $d=3$, calculated using the truncated Fourier-Legendre series in Eq. (\ref{eq:pz-Pn-3d}) for 
$N_c=20$.}
\label{fig:pz3d} 
\end{figure}

In Fig. (\ref{fig:pz2d}) and Fig. (\ref{fig:pz3d}) we compare $p(s)$ recovered from the moments, using 
the formulas in Eq. (\ref{eq:pz-Pn-2d}) and Eq. (\ref{eq:pz-Pn-3d}), with those obtained from simulations.  
\graphicspath{{figures/}}
\graphicspath{{figures/}}
For $d=2$ it is found that the distribution changes from concave to convex at 
$$
\alpha_c = 2.98708\dots
$$
and for $d=3$ at 
$$
\alpha_c = 1.86935\dots,
$$
indicating the decrease of $\alpha_c$ with increasing dimensionality.

\section{Thermal fluctuations}

To incorporate thermal fluctuations into the current theoretical framework, we take advantage of 
the fact that distributions of particles in a harmonic trap subject to two or more independent 
random processes can be represented via a convolution relation \cite{book08,book19,Frydel23b}. 
For a system with one-dimensional geometry (resulting from a potential $u = \frac{1}{2} Kx^2$), 
subject to active motion plus thermal fluctuations, this results in 
\be
p_T(z) =  
 \int_{-1}^{1} dz' \, p(z') p_{eq}(z'-z), 
\label{eq:pT}
\ee
where $p_T(z)$ is the stationary distribution in presence of thermal fluctuations, 
$p_{eq} =    e^{-z^2/2B}  / \sqrt{2\pi B} $ is the Boltzmann distribution in dimensionless units, and 
$
B =   {\mu K}  D /  {v_0^2 }
$ 
is the dimensionless thermal diffusion constant.  The convolution in Eq. (\ref{eq:pT}) amounts to 
Gaussian smearing of $p(z)$.  It needs to be stressed that the convolution construction is not 
a general result of active particles.  It is valid only for a harmonic potential and does not work 
for other types of confinement.

Using Eq. (\ref{eq:pT}), the moments of the distribution $p_T$ can be expressed as 
\be
\langle z^{2n} \rangle_T =   \int_{-\infty}^{\infty} dz\, z^{2n} \int_{-1}^{1} dz' \, p(z') p_{eq}(z-z'). 
\label{eq:z2nT0}
\ee
Then using $z^{2n} = \left[ (z-z') + z' \right]^{2n}$ together with binomial expansion, 
Eq. (\ref{eq:z2nT0}) yields 
\be
\langle z^{2n} \rangle_T    =    \sum _{k=0}^{n} \frac{ (2 n)!  }{(2 k)! (2 n-2 k)!} \langle z^{2 n-2 k}\rangle  \langle z^{2 k} \rangle_{eq}. 
\label{eq:z2nTA}
\ee
Since $\langle z^{2 k} \rangle_{eq}$ can be calculated using the Boltzmann distribution, this leads to 
\be
\langle z^{2n} \rangle_T    =    \sum _{k=0}^{n} \frac{  (2 n)!  B^k }{ 2^{k}  k! (2 n-2 k)!} \langle z^{2 n-2 k}\rangle.  
\label{eq:z2nT00}
\ee
The first two even moments are listed in Table (\ref{table3}) for different dimensions.  The effect of temperature, $B>0$, 
is to increase the value of all moments, corresponding to the spreading of the distributions.  
\begin{table}[h!]
\centering
 \begin{tabular}{ l l l } 
  \hline
\multicolumn{2}{c}{$d=2$}\\
 \hline
 & \\[-1ex]
   ~~~  {$ \langle z^2\rangle_T = \frac{1}{2} \frac{1}{1+\alpha}  +  B$ }  \\ [1.ex] 
   ~~~  {$ \langle z^4\rangle_T = \frac{3}{8} \frac{3 + 4 \alpha}{ (1 + \alpha) (3 + \alpha ) (1 + 2\alpha)}  +   \frac{3B}{1+\alpha}  +  3 B^2 $}    \\ [1.ex] 
   ~~~  \\ [1.ex] 
\hline
\multicolumn{2}{c}{$d=3$}\\
 \hline
& \\[-1ex]
   ~~~  {$ \langle z^2\rangle_T = \frac{1}{3} \frac{1}{1+2\alpha} +  B$ }   \\ [1.ex] 
   ~~~  {$ \langle z^4\rangle_T = \frac{1}{5} \frac{3 + 5 \alpha}{ (1 + 2\alpha) (3 + 2\alpha ) (1 + 3\alpha)}  +  \frac{2 B}{1 + 2 \alpha}  +  3 B^2$}    \\ [1.ex] 
\hline
\end{tabular}
\caption{Moments of a stationary distribution $p_T$ for ABP particles in a potential $u=Kx^2/2$ obtained 
from Eq. (\ref{eq:z2nT00}).  }
\label{table3}
\end{table}

\begin{figure}[hhhh] 
 \begin{center}
 \begin{tabular}{rrrr}
\hspace{-0.3cm}\includegraphics[height=0.15\textwidth,width=0.17\textwidth]{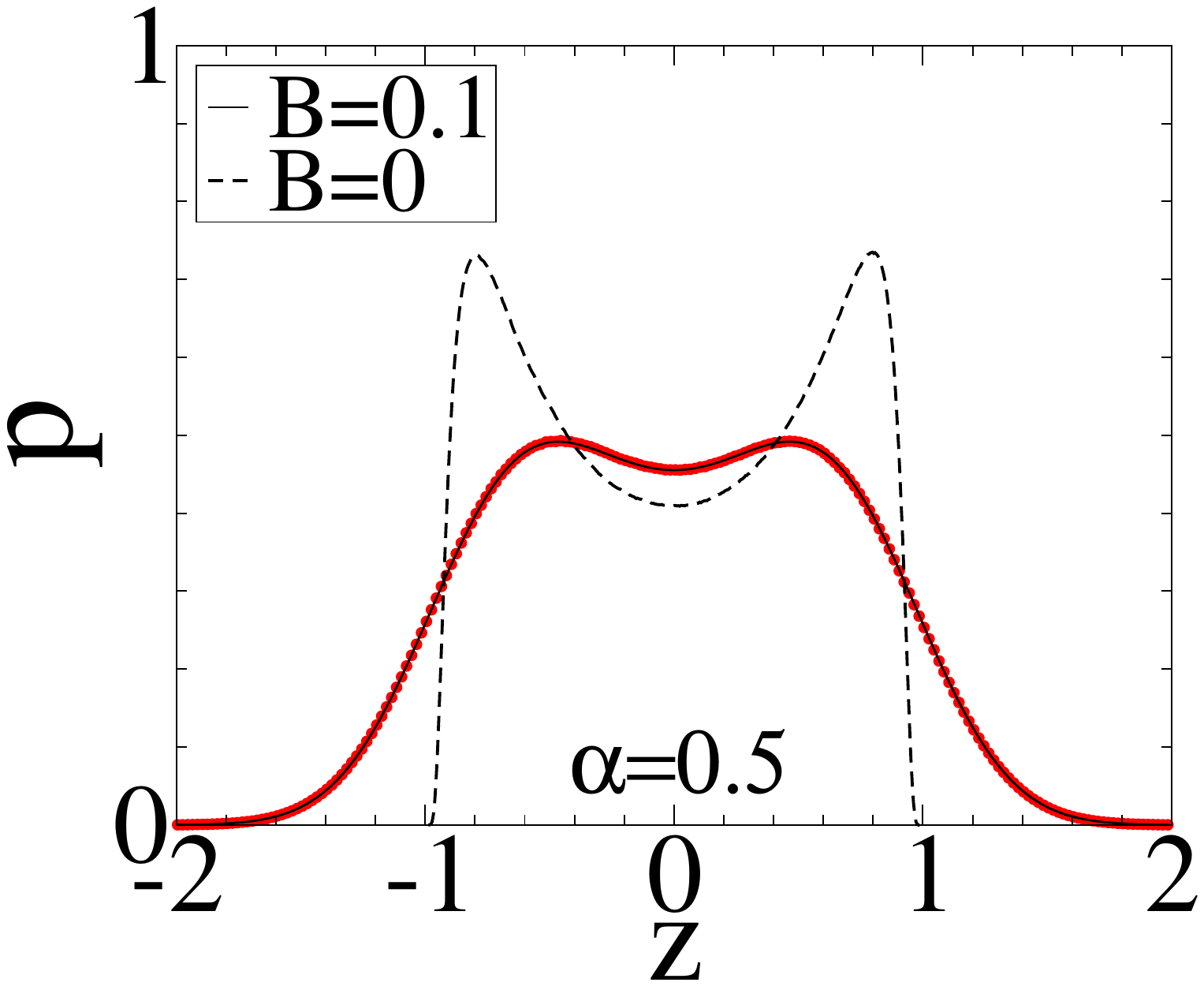} &
\hspace{-0.3cm}\includegraphics[height=0.15\textwidth,width=0.17\textwidth]{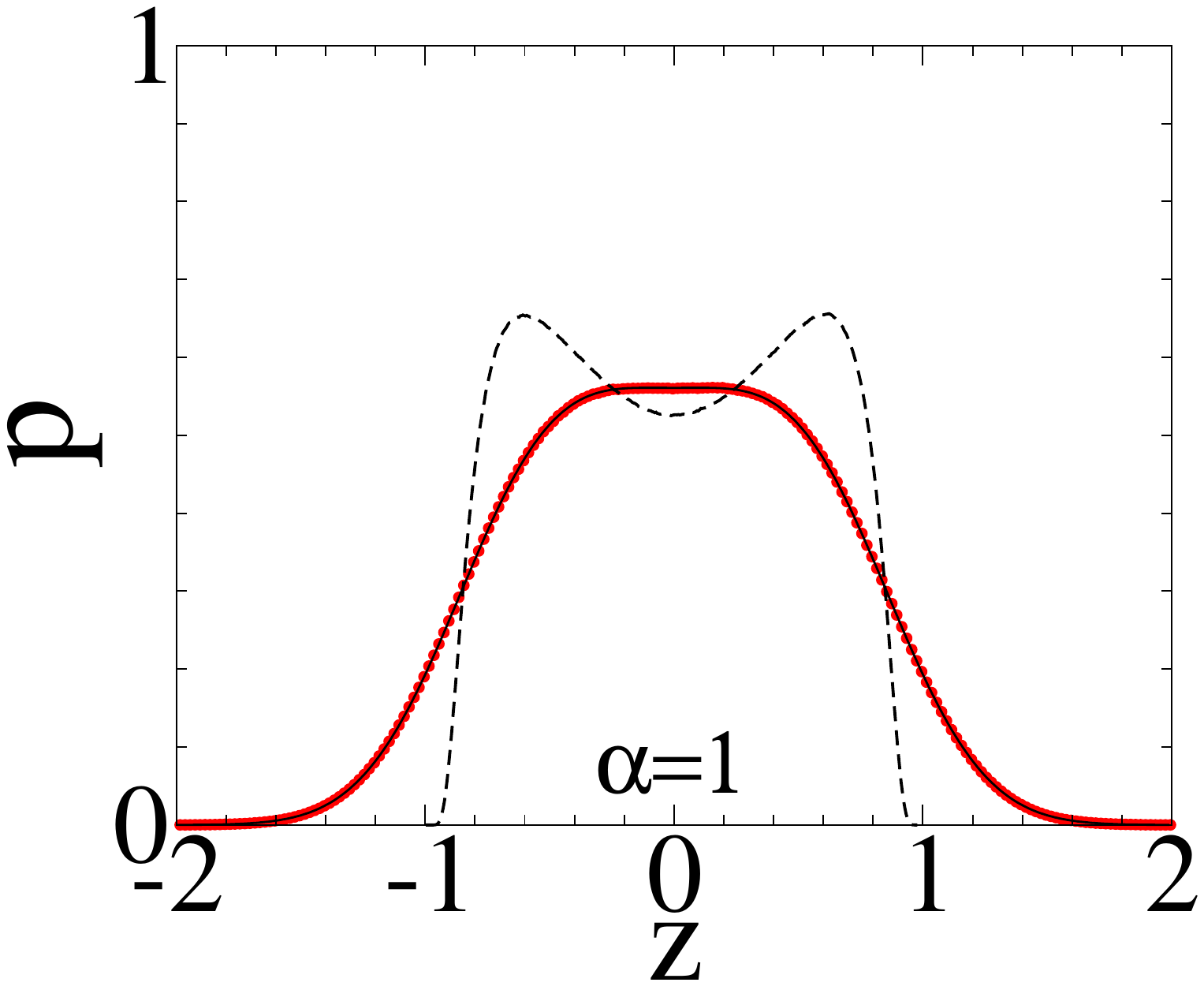} \\
\hspace{-0.2cm}\includegraphics[height=0.15\textwidth,width=0.17\textwidth]{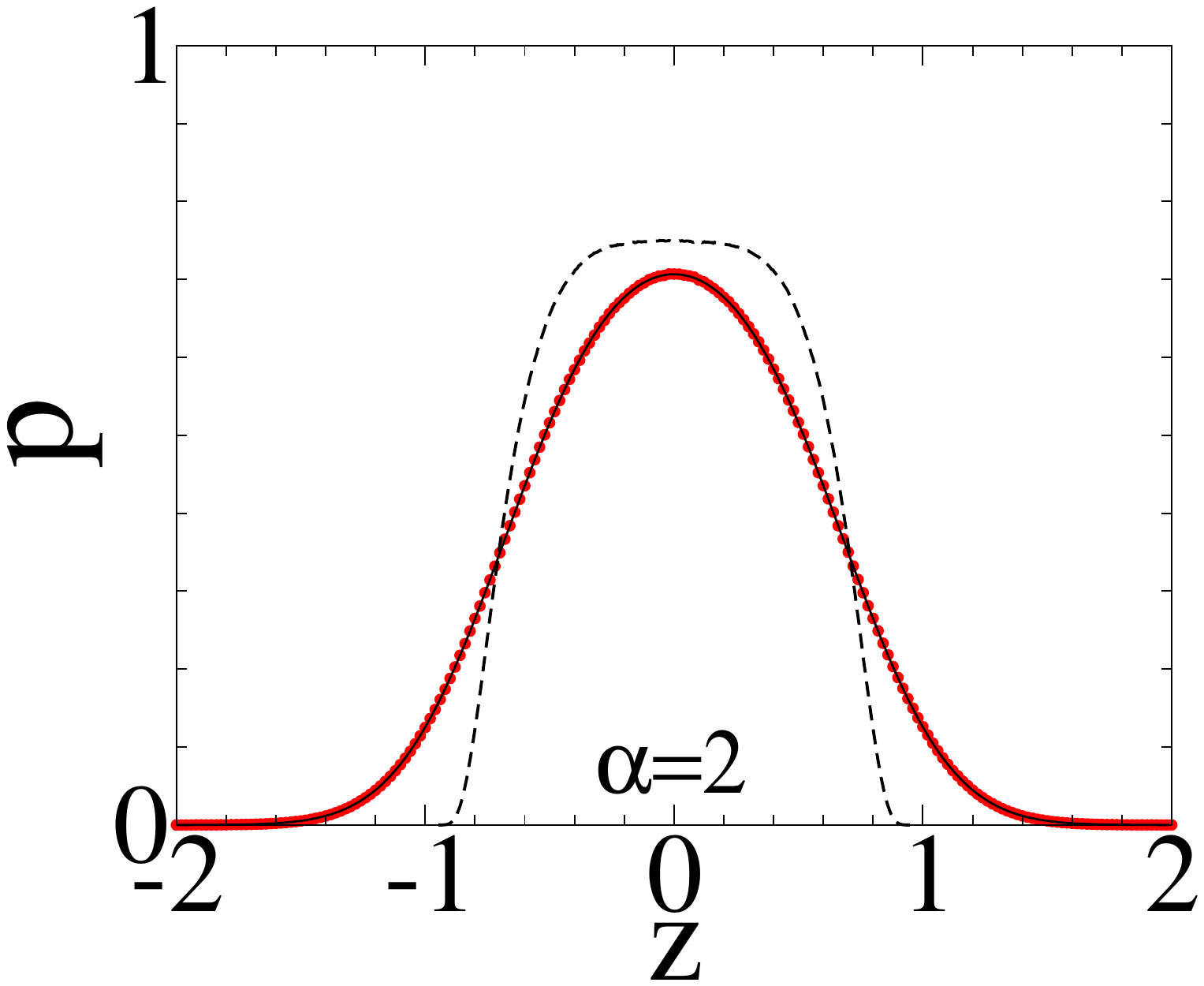} &
\hspace{-0.2cm}\includegraphics[height=0.15\textwidth,width=0.17\textwidth]{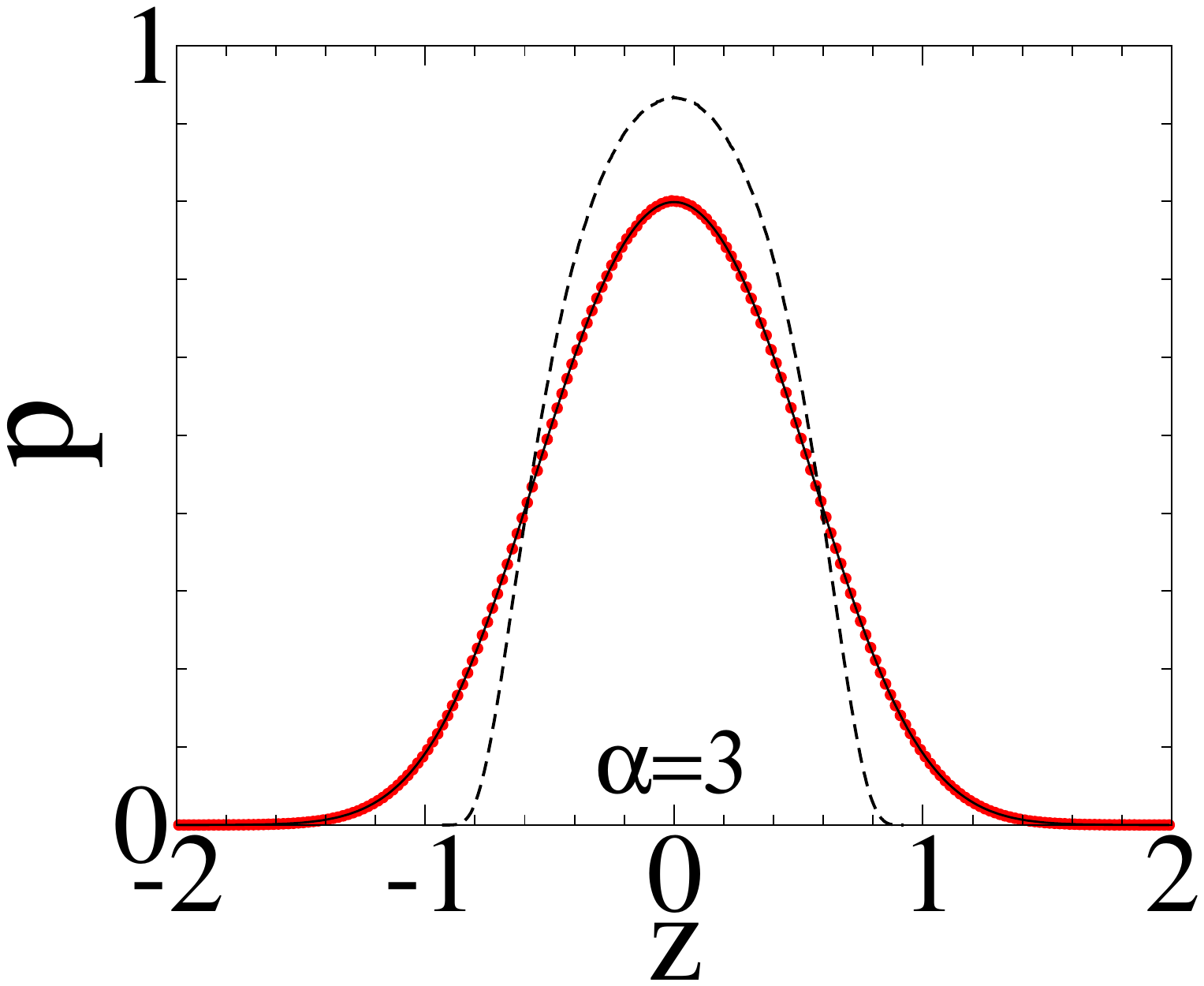} 
 \end{tabular}
 \end{center} 
\caption{Distributions for a finite temperature $p_T$, for $d=2$, calculated using the truncated Fourier-Legendre 
series $p =   \sum_{n=0}^{N_c} a_n P_{2n} (z)$ convoluted according to Eq. (\ref{eq:pT}), for $B=0.1$, 
$N_c=20$, and $a_n$ defined in Eq. (\ref{eq:an}).  
Red circles represent simulation results for $B=0.1$.}
\label{fig:pTz2d} 
\end{figure}

\begin{figure}[hhhh] 
 \begin{center}
 \begin{tabular}{rrrr}
\hspace{-0.3cm}\includegraphics[height=0.15\textwidth,width=0.17\textwidth]{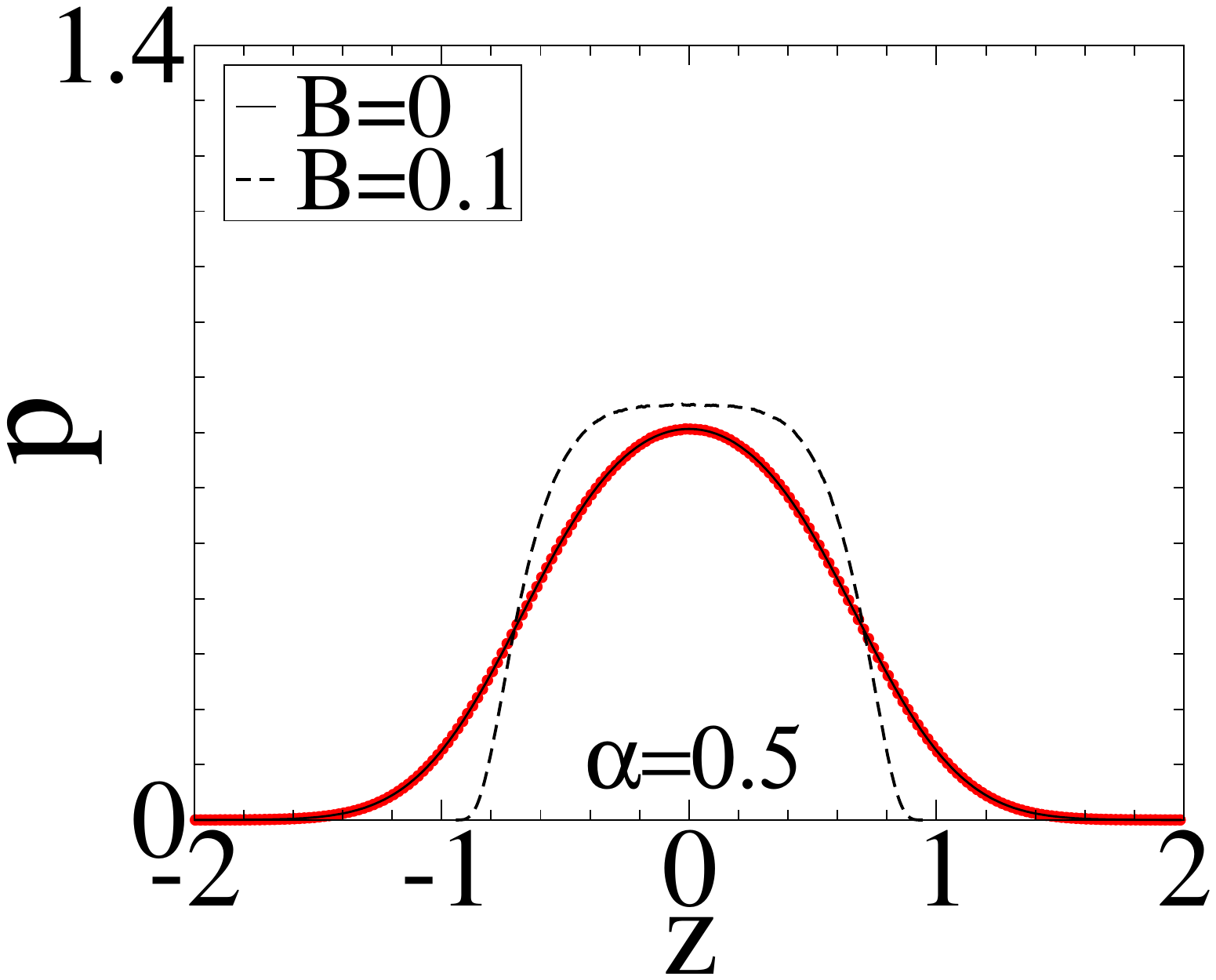} &
\hspace{-0.3cm}\includegraphics[height=0.15\textwidth,width=0.17\textwidth]{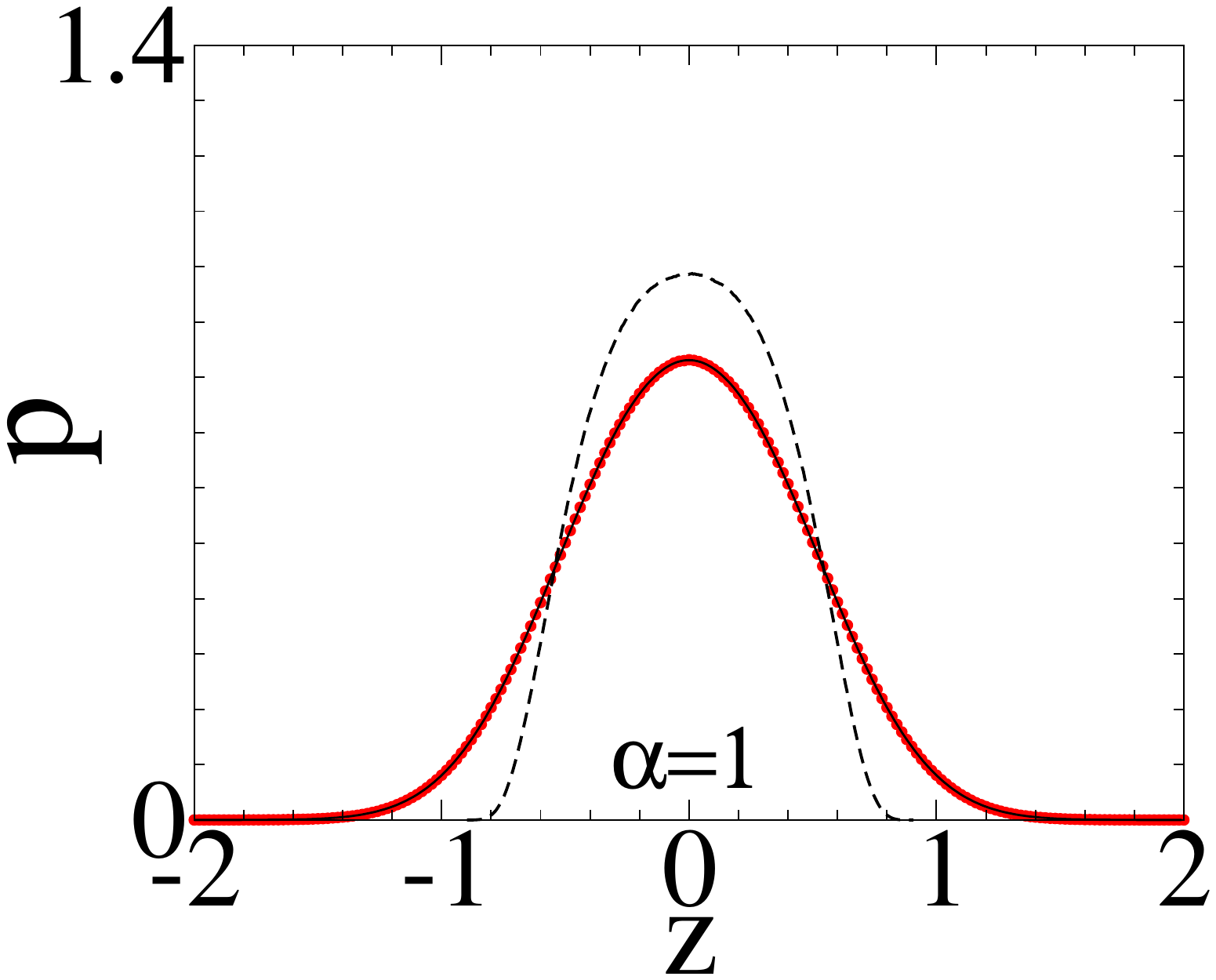} \\
\hspace{-0.2cm}\includegraphics[height=0.15\textwidth,width=0.17\textwidth]{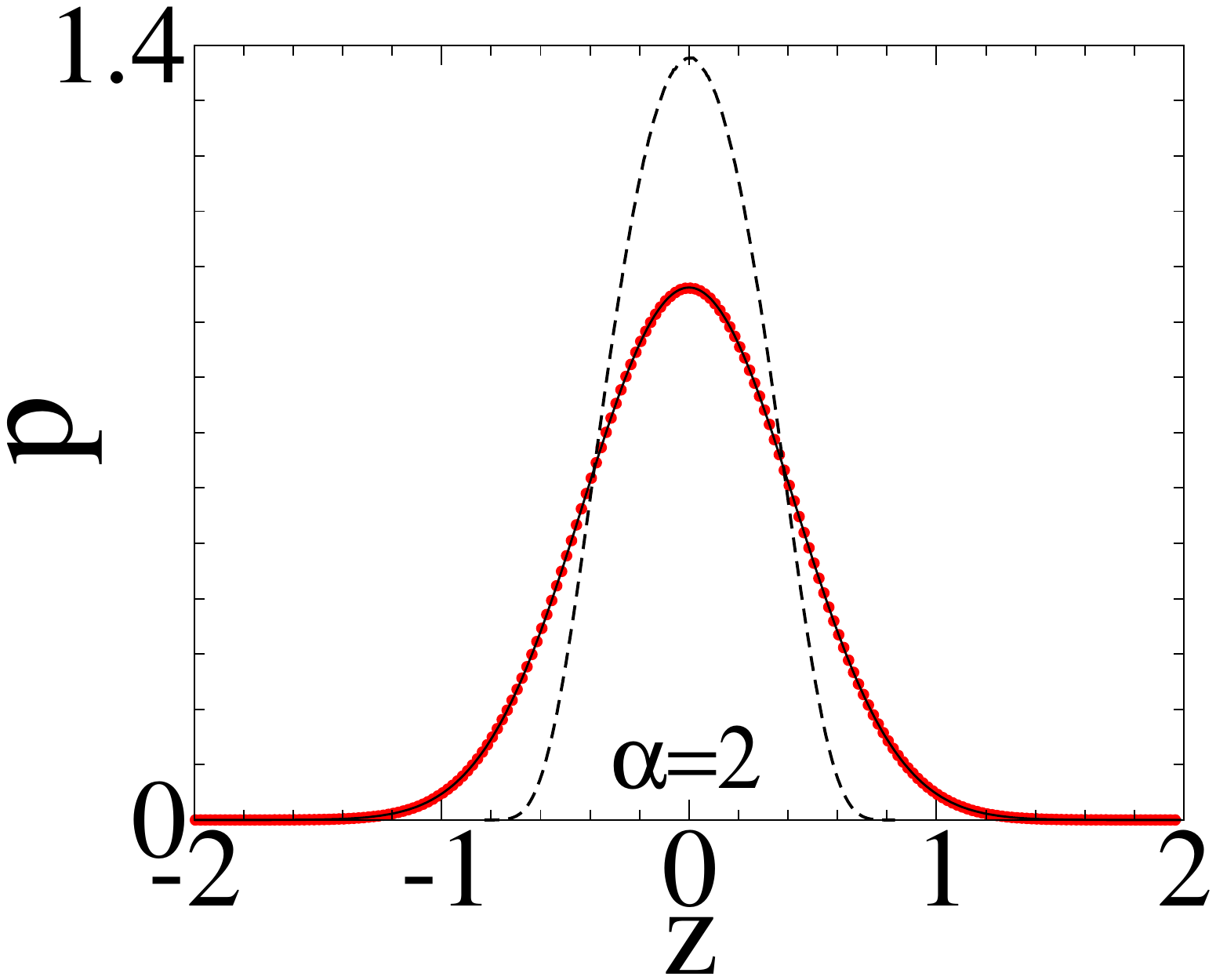} &
\hspace{-0.2cm}\includegraphics[height=0.15\textwidth,width=0.17\textwidth]{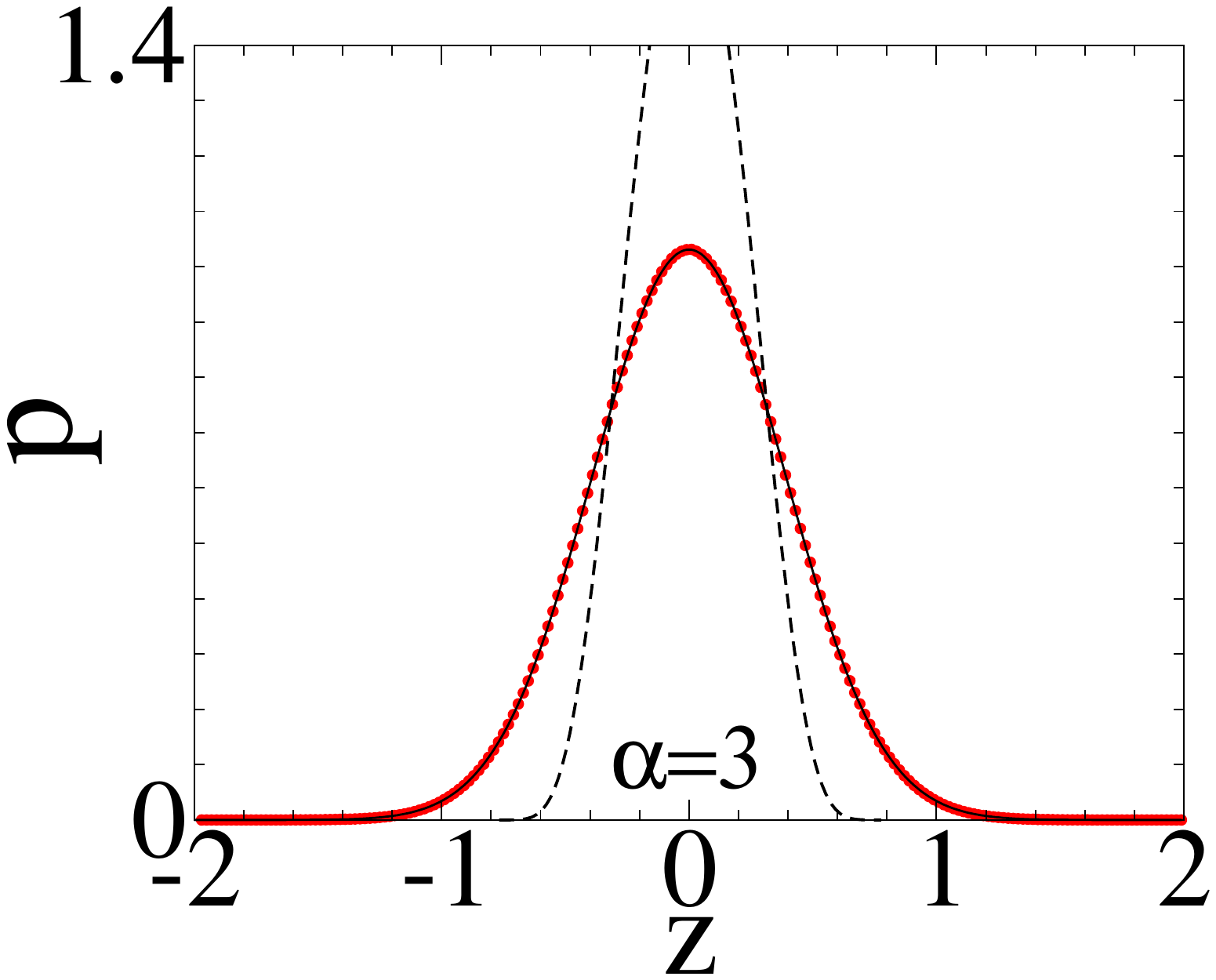} 
 \end{tabular}
 \end{center} 
\caption{Distributions for a finite temperature $p_T$, for $d=3$, calculated using the truncated Fourier-Legendre 
series $p =   \sum_{n=0}^{N_c} a_n P_{2n} (z)$ convoluted according to Eq. (\ref{eq:pT}), for $B=0.1$, 
$N_c=20$, and $a_n$ defined in Eq. (\ref{eq:an}).  Red circles represent simulation results for $B=0.1$.}
\label{fig:pTz3d} 
\end{figure}

Fig. (\ref{fig:pTz2d}) and Fig. (\ref{fig:pTz3d}) show distributions for a finite temperature corresponding 
to $B=0.1$ and for a harmonic potential in a single direction $u=\frac{1}{2} Kx^2$.   
{The distributions for $B>0$ are obtained using the convolution formula in Eq. (\ref{eq:pT}), which amounts
to Gaussian smearing of a distribution at zero temperature.  The smearing effect of the procedure is visible in 
Fig. (\ref{fig:pTz2d}) and Fig. (\ref{fig:pTz3d}), where distributions for a finite $B$ are compared with those 
for $B=0$.  
}

{The value $B=0.1$ was selected for illustrative purposes.  Tthe effect of temperature for this $B$ is sufficiently
strong to be observed as a deviation from the distribution for $B=0$.  At the same time, deviations are 
not strong enough to result in a Gaussian distribution. }
\graphicspath{{figures/}}
\graphicspath{{figures/}}

\section{resetting-ABP model}

The method developed in this work based on the transformation of a stationary FP equation into a recurrence relation 
for generating moments constitutes a general procedure for treating active particles in a harmonic confinement.  

In this section, we show how this method can be applied to the resetting-ABP model 
that combines the RTP and ABP dynamics \cite{Basu20b,Evans20}.   
The stationary FP equation for this system in a linear harmonic potential and in reduced units is 
\ba
0 &=&  \frac{\partial}{\partial z} \left[ \left( z  -   \cos\theta \right) \rho \right]   \nonumber\\ 
&+&  \alpha  \left[ \frac{\partial^2 \rho}{\partial\theta^2}   +   (d-2)  \cot\theta  \frac{\partial \rho}{\partial\theta} \right]   
- \alpha_R \left[ \rho - \frac{p}{ C} \right], 
\ea
where $\alpha_R=\frac{1}{\,u K \tau}$ is the dimensionless rate of orientational change of RTP particles, $\tau$ is 
the persistence time associated with that motion, $C = \int_{0}^{\pi} d\theta\, \sin^{d-2}\theta$, and 
$p = \int_{0}^{\pi} d\theta\,\rho(z,\theta) \sin^{d-2}\theta$ is the marginal distribution.  The FP equation transforms 
into the following recurrence relation:  
\ba
  A_{l,m}   &=&         \left[  \frac{\alpha m(m-1) }{l     +     \alpha  m^2    +    \alpha m (d-2)   +    \alpha_R} \right]  A_{l,m-2}   \nonumber\\
  &+&   \left[ \frac{l}{l       +     \alpha m^2    +    \alpha m (d-2)   +    \alpha_R}   \right]  A_{l-1,m+1}  \nonumber\\ 
  &+&      \left[  \frac{\alpha_R}{l      +     \alpha  m^2    +    \alpha m (d-2)   +    \alpha_R}  \right]   A_{l,0} A_{0,m}.
\label{eq:Alm-rABP}
\ea
For $\alpha_R=0$ the relation reduces to that in Eq. (\ref{eq:Alm}).

The terms $A_{l,m}$ are generated directly from Eq. (\ref{eq:Alm-rABP}).  Details are provided in Appendix (\ref{sec:app1}).      
The moments generated from the recurrence relation, as defined by Eq. (\ref{eq:A2n0}), are listed in Table (\ref{table1}) 
for a dimension $d=2$ and $d=3$.  
\begin{table}[h!]
\centering
 \begin{tabular}{ l l l } 
  \hline
\multicolumn{2}{c}{$d=2$}\\
 \hline
 & \\[-1ex]
   ~~~  {$ \langle z^2\rangle = \frac{1}{2} \frac{1}{1+\alpha + \alpha_R}$ }   \\ [1.ex] 
   ~~~  {$ \langle z^4\rangle = \frac{3}{4} \frac{3 + 4 \alpha + \alpha_R}{ (1 + \alpha + \alpha_R) (3 + \alpha + \alpha_R) (2 + 4\alpha + \alpha_R)}$}    \\ [1.ex] 
\hline
\multicolumn{2}{c}{$d=3$}\\
 \hline
& \\[-1ex]
   ~~~  {$ \langle z^2\rangle = \frac{1}{3} \frac{1}{1+2\alpha+\alpha_R}$ }   \\ [1.ex] 
   ~~~  {$ \langle z^4\rangle = \frac{1}{15} \frac{18 + 30 \alpha + 5\alpha_R}{ (1 + 2\alpha + \alpha_R) (3 + 2\alpha +\alpha_R) (2 + 6\alpha + \alpha_R)}$}    \\ [1.ex] 
\hline
\end{tabular}
\caption{Moments of a stationary distribution generated from the recurrence relation in Eq. (\ref{eq:Alm-rABP}).  }
\label{table1}
\end{table}

\section{Conclusion}

The current article provides a theoretical framework for treating active particles in a harmonic well
based on a simple transformation of a stationary Fokker-Planck equation into a recurrence relation.  
The method applies to any dimension and type of active dynamics and, as such, provides a 
general framework for treating active oscillators.  

The methodology in this work takes advantage of the convolution construction of distributions in 
a harmonic well to incorporate thermal fluctuations in a systematic way.  The current work, furthermore, 
provides a relation in Eq. (\ref{eq:r2n}) between moments of a distribution for a linear and isotropic harmonic 
potentials.  This relation was proven before for RTP particles.  In this work it is shown that the same relation 
applies to ABP particles, which shows Eq. (\ref{eq:r2n}) to be a general result valid for any system of active
particles.  

{It remains to be seen if the method used in this work can be extended to other types of systems.  
Some suggestions for future direction are, for example, to see if the current method is applicable to active 
particles in an underdamped regime.  There is also a question whether the method applies to other types of 
confinements.  Lastly, it remains to be seen if the method can shed any light or offer guidelines
for understanding and treating many particle systems \cite{Wang21,Vinze21}.   }



\section{DATA AVAILABILITY}
The data that support the findings of this study are available from the corresponding author upon 
reasonable request.

\appendix
\section{Solving recurrence relation $A_{l,m}$}
\label{sec:app1}

\begin{figure}[hhhh] 
 \begin{center}
 \begin{tabular}{rrrr}
\includegraphics[height=0.2\textwidth,width=0.2\textwidth]{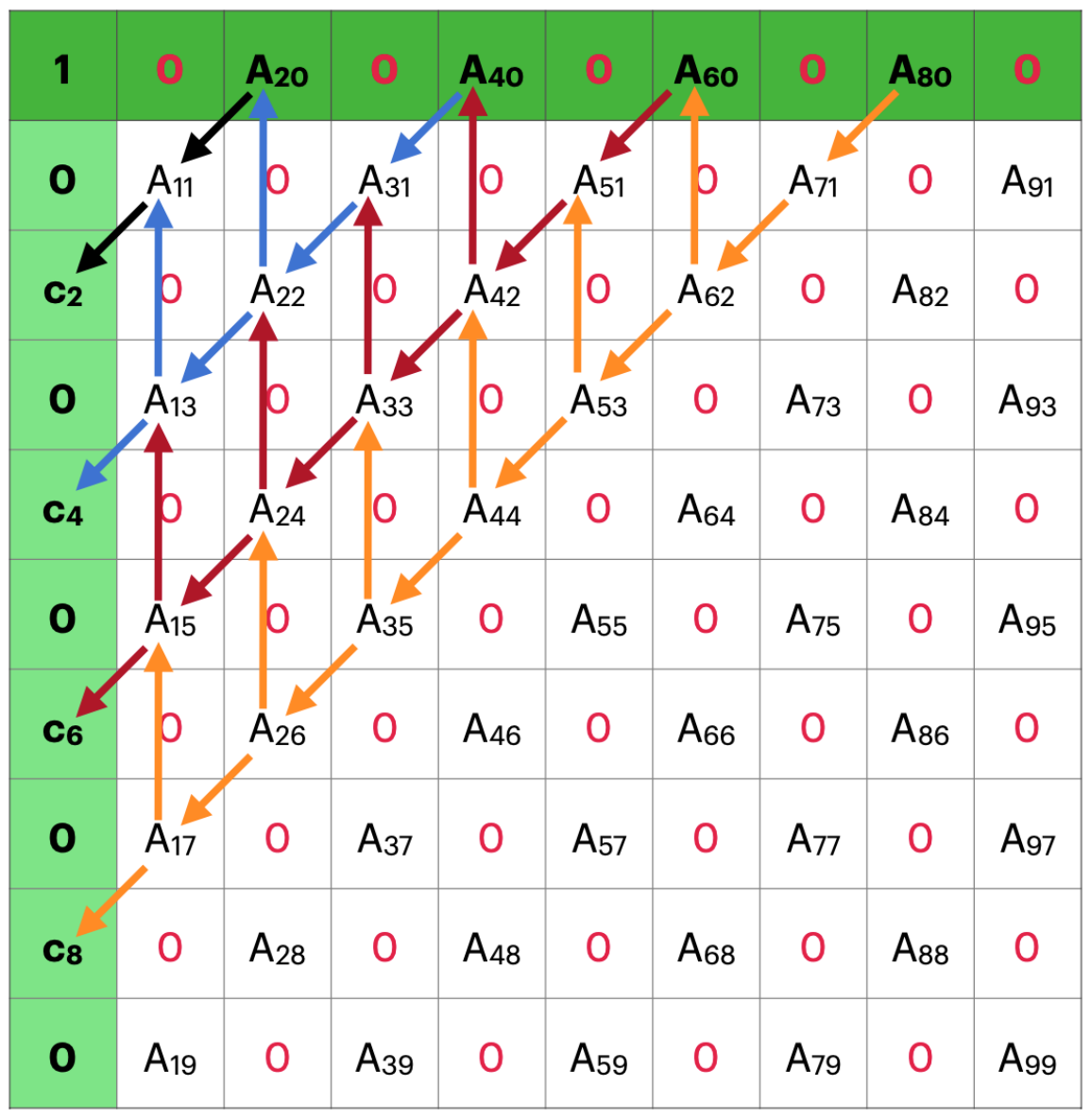} &&
 \end{tabular}
 \end{center} 
\caption{ The coefficient table with arrows indicating the interdependence between different terms.}
\label{fig:Alm} 
\end{figure}

This section provides the details of how to generate coefficients $A_{l,m}$ from the 
recurrence relation in Eq. (\ref{eq:Alm}) that we reproduce below 
\ba
A_{l,m}   &=&   \left[ \frac{ \alpha m^2 - \alpha m   }{  l +  \alpha m^2 + (d-2)\alpha m  }  \right]  A_{l,m-2} \nonumber\\ 
&+& \left[ \frac{l}{  l +  \alpha m^2 + (d-2) \alpha m } \right]  A_{l-1,m+1}.  
\label{eq:app1a}
\ea
The initial condition of the recurrence relation are the coefficients $A_{0,m}$ given by 
\be
A_{0,m}  = c_m = 
\begin{dcases*}
 \frac{\Gamma \left(\frac{d}{2}\right) \Gamma \left(\frac{m}{2}+\frac{1}{2}\right)}{\sqrt{\pi } \Gamma \left(\frac{d}{2}+\frac{m}{2}\right)}, & \text{$m = $ even}\\
0, & \text{$m = $ odd} 
\end{dcases*}
\label{eq:app1b}
\ee
It is noted that 
$$
A_{l,m}=0, ~~~~ \text{$l+m = $ odd}. 
$$
Using the recurrence relation in Eq. (\ref{eq:app1a}) we find
$$
A_{2,0} = A_{1,1} =  \frac{ c_{2} }{  1 +  \alpha + (d-2) \alpha }, 
$$
where $c_{2n}$ is defined in Eq. (\ref{eq:app1b}).  
The chain of relations in Fig. (\ref{fig:Alm}) indicates that $A_{2,0}$ is related to $A_{1,1}$ which in turn is related to $c_2$.  

To calculate $A_{4,0}$ we go through the following sequence of terms:   
$A_{4,0}  \to A_{3,1} \to A_{2,2} \to A_{2,0} + A_{1,3} \to A_{2,0} + A_{1,1} + c_{4}$.  
This means that $A_{4,0}$ is represented in terms of $c_{4}$ and the coefficients $A_{2,0}$ and $A_{1,1}$
calculated previously.  
\graphicspath{{figures/}}

To generalize this procedure to an arbitrary $A_{2n,0}$ using Mathematica, we use the
following code line:
{\small
\ba
&&\text{Table} \Big[ l = k;\, m = 2n - k; \nonumber\\ 
&& 
A[l][m] = \frac{\alpha m^2 - \alpha m} {l + \alpha m^2 + (d-2)\alpha m} A[l][m-2]   \nonumber\\ 
&& + \frac{l}{l+a m^2 + (d-2)\alpha m} A[l-1][m+1],
\nonumber\\ &&
\{n,1,2N_c\},\{k,1,2n-1\} \Big]
\ea
}

\noindent
keeping in mind that $A_{2n-1,1}=A_{2n,0}$.

\section{Isotropic harmonic potential}
\label{sec:app3}

To confirm the relation in Eq. (\ref{eq:r2n}) for active particles, we start with the stationary FP equation for particles in an 
isotropic harmonic trap
$$
0  =  \bnabla \cdot \left[ \left( \mu K {\bf r} - v_0 {\bf n} \right) \rho \right]   + \hat L\, \rho, 
$$

Because $\rho$ depends on a relative orientation of the vectors ${\bf r}$ and ${\bf n}$, we are allowed to fix 
${\bf n}$ along the $x$-axis, ${\bf n} = {\bf e}_x$.  The FP equation in this case becomes 
\be
0 =  \mu K  {\bf r}  \cdot \bnabla \rho  +  \mu K \rho (\bnabla \cdot {\bf r}) - v_0  \frac{\partial \rho}{\partial x}  
+  \hat L\, \rho.  
\ee
In polar and spherical coordinates 
$\frac{\partial \rho}{\partial x} = \cos\theta \frac{\partial\rho}{\partial r} -   \frac{\sin\theta}{r} \frac{\partial \rho}{\partial \theta}$,
${\bf r} \cdot \bnabla \rho  = r \frac{\partial \rho}{\partial r}$, and $\bnabla \cdot {\bf r} = d$.  Together with 
$\hat L\rho$ given in Eq. (\ref{eq:FPS}), the FP equation in reduced units $s = \mu K r/v_0$ becomes 
\ba
0 &=&  \frac{\partial}{\partial s} \left[ \left( s  -   \cos\theta \right) \rho \right] 
+  (d-1) \rho  +    \frac{\sin\theta}{s} \frac{\partial \rho}{\partial \theta} \nonumber\\ 
&+&  \alpha  \left[ \frac{\partial^2 \rho}{\partial\theta^2}   +   (d-2)  \cot\theta  \frac{\partial \rho}{\partial\theta} \right].  
\label{eq:FPS-R}
\ea

The recurrence relation corresponding to the above FP equation is  
\ba
  B_{l,m}   &=&         \left[  \frac{\alpha m(m-1) }{l     +     \alpha  m^2    +    \alpha m (d-2)  } \right]  B_{l,m-2}   \nonumber\\
  &+&   \left[ \frac{l - m}{l       +     \alpha m^2    +    \alpha m (d-2)   }   \right]  B_{l-1,m+1}  \nonumber\\ 
  &+&   \left[ \frac{m}{l       +     \alpha m^2    +    \alpha m (d-2)   }   \right]  B_{l-1,m-1}.
\label{eq:Bn}
\ea
where $B_{2n,0} = \langle s^{2n} \rangle$ and the only known coefficient in this case is $B_{0,0}=1$, which is enough to calculate 
$B_{2n,0}$.  


The computed coefficients $B_{2n,0}$ from the recurrence relation in Eq. (\ref{eq:Bn}) 
are found to satisfy the following relation 
$$
B_{2n,0}  =  \frac{\Gamma \left(\frac{1}{2}\right) \Gamma \left(\frac{d}{2}+n\right)}{\Gamma \left(\frac{d}{2}\right) \Gamma \left(n+\frac{1}{2}\right)} A_{2n,0}, 
$$
where $A_{2n,0}$ are the coefficients in Eq. (\ref{eq:Alm}) .  
It is concluded, therefore, that the relation in Eq. (\ref{eq:r2n}) is correct.

\section{Recovering $p(s)$ from the moments for $d=2$}
\label{sec:app4}

For $d=2$, the normalized distribution is $2\pi s p(s)$.  This results in an odd function on $[-1,1]$, 
however, we have only even moments available to us.  To resolve this problem, we use the change 
of variables $t= s^2$, and then transform the integrated distribution as 
$$
A_{2n,0}   =   2\pi \int_{0}^1 dr\, s p(s) s^{2n}  =  \pi \int_{0}^1 dt\,  p(t) t^{n}. 
$$
The distribution $\pi p(t)$ contains both even and odd terms.   
And since $\langle t^{n}\rangle = A_{2n,0}$, we can determine all the moments.    

Because Legendre polynomials are defined on $[-1,1]$ but the distribution of interest is defined on $[0,1]$, 
we need to shift the Legendre polynomials as 
\be
\pi p(t) = \sum_{m=0}^{\infty} a_m P_m(2t-1)
\ee
where the coefficients are defined as
\be
a_m = (2m + 1)  \int_{0}^{1} dt\, P_m \left( 2t - 1 \right)  \pi p(t).  
\ee
Given the explicit formula for Legendre polynomials 
\be
P_m = 2^m \sum_{k=0}^m t^k  {m \choose k}  {\frac{m+k-1}{2} \choose m},
\ee
we find the coefficients to be 
\be
a_m = 2^m (2m + 1)  \sum_{k=0}^m  {m \choose k}  {\frac{m+k-1}{2} \choose m}   \langle (2t-1)^k \rangle.  
\ee

After changing variables back to $s$ we get 
\be
p(s) = \frac{1}{\pi} \sum_{m=0}^{\infty} a_m P_m(2s^2-1), 
\ee
with the coefficients given by 
\be
a_m = 2^m (2m + 1)  \sum_{k=0}^m  {m \choose k}  {\frac{m+k-1}{2} \choose m}  \sum_{p=0}^k { k \choose p } 2^{p} (-1)^{k-p}   \langle s^{2p} \rangle.
\ee




\begin{thebibliography}{99}


\bibitem{Frydel23b}
D. Frydel, 
"{\sl Run-and-tumble oscillator:  moment analysis of stationary distributions}", 
Phys. Fluids, {\bf 35}, 101905 (2023).  

\bibitem{Cates08}
J. Tailleur and M. E. Cates, 
{\sl Statistical Mechanics of Interacting Run-and-Tumble Bacteria}, 
Phys. Rev. Lett {\bf 100}, 218103 (2008).

\bibitem{Cates09}
J. Tailleur and M. E. Cates,
{\sl Sedimentation, trapping, and rectification of dilute bacteria}, 
Europhys. Lett. {\bf 86}, 60002 (2009).

\bibitem{Dhar19}
A. Dhar, A. Kundu, S. N. Majumdar, S. Sabhapandit, and G. Schehr
{\sl Run-and-tumble particle in one-dimensional confining potentials: Steady-state, relaxation, and first-passage properties}
Phys. Rev. E {\bf 99}, 032132 (2019). 

\bibitem{Frydel22c}
D. Frydel,
{\sl Positing the problem of stationary distributions of active particles as third-order differential equation}, 
Phys. Rev. E {\bf 106}, 024121 (2022). 

\bibitem{Scher22}
N. R. Smith, P. Le Doussal, S. N. Majumdar, G. Schehr, 
{\sl Exact position distribution of a harmonically confined run-and-tumble particle in two dimensions}, 
Phys. Rev. E {\bf 106}, 054133 (2022).

\bibitem{Frydel23}
D. Frydel,
{\sl Entropy production of active particles formulated for underdamped dynamics}, 
Phys. Rev. E {\bf 107}, 014604 (2023).





\bibitem{Brady16}
 S. C. Takatori, R. De Dier, J. Vermant, and J. F. Brady, 
 {\sl Acoustic trapping of active matter},
 Nat. Commun. {\bf 7}, 10694 (2016).

\bibitem{Lowen22}
I. Buttinoni, L. Caprini, L. Alvarez , F. J. Schwarzendahl, and H. L\"owen, 
{\sl Active colloids in harmonic optical potentials},
EPL {\bf 140},  27001 (2022). 





\bibitem{Szamel14}
G. Szamel, 
{\sl Self-propelled particle in an external potential: Existence of an effective temperature},
Phys, Rev. E {\bf 90}, 012111 (2014).  

\bibitem{Fodor16}
\'E. Fodor, C. Nardini, M. E. Cates, J. Tailleur, P. Visco, and F. van Wijland 
{\sl How Far from Equilibrium Is Active Matter?},  
Phys. Rev. Lett. {\bf 117}, 038103 (2016).

\bibitem{Martin21}
D. Martin, J. O’Byrne, M. E. Cates, E. Fodor, C. Nardini, J. Tailleur, and F. van Wijland, 
{\sl Statistical mechanics of active Ornstein-Uhlenbeck particles}, 
Phys. Rev.  E {\bf 103}, 032607 (2021).  









\bibitem{Basu20}
U. Basu, S.N. Majumdar, A. Rosso, S. Sabhapandit and G. Schehr, 
{\sl Exact stationary state of a run-and-tumble particle with three internal states in a harmonic trap}, 
J. Phys. A: Math. Theor. {\bf 53}, 09LT01 (2020).





\bibitem{Carpini22}
L. Caprini, A. R. Sprenger, H. L\"owen, R Wittmann, 
{\sl The parental active model: A unifying stochastic description of self-propulsion},  
J. Chem. Phys. {\bf 156}, 071102 (2022).



\bibitem{Farago22}
N. R. Smith and O. Farago, 
{\sl Nonequilibrium steady state for harmonically confined active particles}, 
Phys. Rev. E {\bf 106}, 054118 (2022).








%
%



%
%































\bibitem{Gupta21}
D. Gupta and D. A. Sivak, 
{\sl Heat fluctuations in a harmonic chain of active particles}, 
Phys.Rev. E {\bf 104}, 024605 (2021).

\bibitem{Kundu21}
P. Singh and A. Kundu, 
{\sl Crossover behaviours exhibited by fluctuations and correlations in a chain of active particles}, 
J. Phys. A: Math. Theor. {\bf 54}, 305001 (2021).  

\bibitem{Basu22}
I. Santra, U. Basu, 
{\sl Activity driven transport in harmonic chains}, 
SciPost Phys. {\bf 13}, 041 (2022). 


%















%
%
%



%

































%
%
%
%







\bibitem{Caprini19}
L. Caprini, U. M. B. Marconi, A. Puglisi, A. Vulpiani, 
{\sl The entropy production of Ornstein-Uhlenbeck active particles: a path integral method for correlations}, 
J. Stat. Mech. 053203, (2019).

\bibitem{Dabelow21}
L. Dabelow, S. Bo, R. Eichhorn,
{\sl How irreversible are steady-state trajectories of a trapped active particle?}, 
J. Stat. Mech. 033216, (2021).

\bibitem{Frydel22a}
D. Frydel,
{\sl Intuitive view of entropy production of ideal run-and-tumble particles}, 
Phys. Rev. E {\bf 105}, 034113 (2022).

\bibitem{Dabelow19}
L Dabelow, S Bo, R Eichhorn, 
{\sl Irreversibility in active matter systems: Fluctuation theorem and mutual information}, 
Phys. Rev. X {\bf 9}, 021009, (2019).   


\bibitem{Pruessner21}
R. Garcia-Millan and G. Pruessner 
{\sl Run-and-tumble motion in a harmonic potential: field theory and entropy production}, 
J. Stat. Mech. 063203 (2021).  





\bibitem{Dhar20}
K. Malakar, A. Das, A. Kundu, K. V. Kumar, and A. Dhar
{\sl Steady state of an active Brownian particle in a two-dimensional harmonic trap}, 
Phys. Rev. E {\bf 101}, 022610 (2020).

\bibitem{Cargalio22}
M. Caraglio, T. Franosch, 
{\sl Analytic solution of an active brownian particle in a harmonic well},
Phys. Rev. Lett. {\bf 129}, 158001, (2022).  





\bibitem{book08}
N. F. Sharpe and R. F. Carter, Genetic Testing (John Wiley \& Sons, Inc., Hoboken, NJ, 2005).
Dimitri P. Bertsekas and John N. Tsitsiklis, {\sl Introduction To Probability}
(Athena Scientific 2008) 2nd Ed. 

\bibitem{book19}
Joseph K. Blitzstein and Jessica Hwang, {\sl Introduction to Probability}
 (Chapman \& Hall/CRC Texts in Statistical Science 2019) 2nd Ed.  




\bibitem{Basu20b}
V. Kumar, O. Sadekar, and U. Basu, 
{\sl Active Brownian motion in two dimensions under stochastic resetting}, 
Phys. Rev. E {\bf 102}, 052129 (2020).

\bibitem{Evans20}
M. R. Evans, S. N. Majumdar and G. Schehr, 
{\sl Stochastic resetting and applications}, 
J. Phys. A: Math. Theor. {\bf 53} 193001 (2020).






\bibitem{Vinze21}
Prathmesh M. Vinze, Akash Choudhary, S. Pushpavanam, 
{\sl Motion of an active particle in a linear concentration gradient Special Collection: Dynamics of Out-of-Equilibrium Soft Materials},
Phys. Fluids {\bf 33}, 032011 (2021).  

\bibitem{Wang21}
B. Deußen, M. Oberlack, Y. Wang 
{\sl Probability theory of active suspensions}
Phys. Fluids {\bf 33}, 061902 (2021)

\end{thebibliography}
\end{document}